\documentclass{aa}  

\usepackage{graphicx}
\usepackage{txfonts}
\usepackage{lipsum}
\usepackage{subcaption}
\usepackage{lscape}
\usepackage{placeins}
\usepackage[percent]{overpic}
\usepackage{multirow}
\usepackage{breakcites}
\usepackage[nopatch=footnote]{microtype}
\usepackage{enumitem}
\usepackage{booktabs}
\usepackage{scalerel}
\usepackage{etoolbox}
\usepackage{siunitx}
\DeclareSIUnit{\magnitude}{mag}
\DeclareSIUnit{\jansky}{Jy}
\DeclareSIUnit{\beam}{beam}
\DeclareSIUnit{\deg}{deg}
\DeclareSIUnit{\asec}{arcsec}
\DeclareSIUnit{\yr}{yr}
\DeclareSIUnit{\msun}{\ensuremath{\mathrm{M}_{\odot}}}
\DeclareSIUnit{\parsec}{pc}
\DeclareSIUnit{\steradian}{sr}
\usepackage{tikz}
\usetikzlibrary{matrix,positioning}

\NewDocumentCommand{\confusionmatrixperc}{O{false}mmmmmm}{
  \pgfmathsetmacro{\maxval}{max(max(#2,#3),max(#4,#5))}
  \pgfmathsetmacro{\tra}{110-100*ln(#2+1)/ln(\maxval+1)}
  \pgfmathsetmacro{\trb}{110-100*ln(#3+1)/ln(\maxval+1)}
  \pgfmathsetmacro{\trc}{110-100*ln(#4+1)/ln(\maxval+1)}
  \pgfmathsetmacro{\trd}{110-100*ln(#5+1)/ln(\maxval+1)}

  \pgfmathsetmacro{\rowA}{#2+#3}
  \pgfmathsetmacro{\rowB}{#4+#5}

  \pgfmathsetmacro{\percAA}{\rowA > 0 ? round(100 * 100 * (#2/\rowA))/100 : 0}%
  \pgfmathsetmacro{\percAB}{\rowA > 0 ? round(10 * 100 * (#3/\rowA))/10 : 0}%
  \pgfmathsetmacro{\percBA}{\rowB > 0 ? round(10 * 100 * (#4/\rowB))/10 : 0}%
  \pgfmathsetmacro{\percBB}{\rowB > 0 ? round(100 * 100 * (#5/\rowB))/100 : 0}%
  
  \begin{tikzpicture}
  \matrix (m) [matrix of nodes,
   nodes={draw, minimum size=1.25cm, inner sep=0mm, align=center},
   column sep=-1.5pt, row sep=-1.5pt, line width=1.5pt, ampersand replacement=\&]
  {
    \ifbool{#1}{
      |[fill=gray!\tra!]| \shortstack{\pgfmathprintnumber[fixed, precision=1]{\percAA}\%\\(\num{#2})} \& |[fill=gray!\trb!]| \shortstack{\pgfmathprintnumber[fixed, precision=1]{\percAB}\%\\(\num{#3})} \\
      |[fill=gray!\trc!]| \shortstack{\pgfmathprintnumber[fixed, precision=1]{\percBA}\%\\(\num{#4})} \& |[fill=gray!\trd!]| \shortstack{\pgfmathprintnumber[fixed, precision=1]{\percBB}\%\\(\num{#5})} \\
    }{
      \num{#2} \& \num{#3} \\
      \num{#4} \& \num{#5} \\
  }
  };

  \node [left=0.35cm of m] (true) {\rotatebox{90}{True class}};
  \node [left=0.1cm of m-1-1] {\rotatebox{90}{#7}};
  \node [left=0.1cm of m-2-1] {\rotatebox{90}{#6}};
  \node [below=0.1cm of m-2-1] {#7};
  \node [below=0.1cm of m-2-2] {#6};
  \node [below=0.35cm of m] (pred) {Predicted class};
  \end{tikzpicture}
}

\usepackage[colorlinks=true, linkcolor=blue, citecolor=blue,
            filecolor=blue, urlcolor=blue]{hyperref}

\defcitealias{2023A&A...679A.101C}{C23}
\defcitealias{2012MNRAS.426.3271M}{M12}
\defcitealias{2018MNRAS.478.3056B}{B18}

\begin{document}

   \title{Evolutionary Map of the Universe (EMU): Selection of active galactic nuclei and star-formation emission with machine-learning techniques}

   \titlerunning{Selection of AGN and SF activity in the EMU-PS with ML}

   \author{R. Carvajal\inst{1, 2, 3}\corrauth{racarvajal@ciencias.ulisboa.pt}
          \and I. Matute\inst{1, 4}
          \and J. Afonso\inst{1, 4}
          \and H. Messias\inst{5, 6}
          \and M. J. I. Brown\inst{7}
          \and D. Farrah\inst{8, 9}
          \and A. M. Hopkins\inst{10, 11}
          \and J. Th.\ van Loon\inst{12}
          \and K. J. Luken\inst{13, 14}
          \and C. M. Pennock\inst{15}
          \and J. Prathap\inst{10, 11, 16}
          \and S. S. Shabala\inst{17}
          \and D. A. Leahy\inst{18}
          \and S. F. Rahman\inst{19, 20}
          \and K. A. Pimbblet\inst{21}
          \and T. Zafar\inst{10}
        }

   \institute{Instituto de Astrof\'{\i}sica e Ci\^{e}ncias do Espa\c{c}o, Universidade de Lisboa, OAL, Tapada da Ajuda, 1349-018 Lisbon, Portugal
        \and Departamento de F\'{\i}sica Te\'{o}rica, At\'{o}mica y \'{O}ptica, Universidad de Valladolid, 47011, Valladolid, Spain
        \and Laboratory for Disruptive Interdisciplinary Science (LaDIS), Universidad de Valladolid, 47011 Valladolid, Spain
        \and Departamento de F\'{\i}sica, Faculdade de Ci\^encias, Universidade de Lisboa, Edif\'{\i}cio C8, Campo Grande, 1749-016 Lisbon, Portugal
        \and Joint ALMA Observatory, Alonso de C\'{o}rdova 3107, Vitacura 763-0355, Santiago, Chile
        \and European Southern Observatory, Alonso de C\'{o}rdova 3107, Vitacura, Casilla 19001, Santiago de Chile, Chile
        \and School of Physics, Monash University, Clayton, VIC 3800, Australia
        \and Department of Physics and Astronomy, University of Hawai`i at M{\=a}noa, 2505 Correa Rd., Honolulu, HI, 96822, USA
        \and Institute for Astronomy, University of Hawai`i,  2680 Woodlawn Dr., Honolulu, HI, 96822, USA
        \and School of Mathematical and Physical Sciences, 12 Wally's Walk, Macquarie University, NSW 2109, Australia
        \and Macquarie University Astrophysics and Space Technologies Research Centre, Macquarie University, Sydney, Australia
        \and Lennard-Jones Laboratories, Keele University, Staffordshire ST5 5BG, UK
        \and School of Science, Western Sydney University, Locked Bag 1797, Penrith, NSW 2751, Australia
        \and CSIRO Data61, P.O. Box 76, Epping, NSW 1710, Australia
        \and Institute for Astronomy, University of Edinburgh, Royal Observatory, Edinburgh, EH9 3HJ, UK
        \and Australia Telescope National Facility, CSIRO, Space and Astronomy, PO Box 1130, Bentley, WA 6102, Australia
        \and School of Natural Sciences, University of Tasmania, Private Bag 37, Hobart, Tasmania 7001, Australia
        \and Department of Physics and Astronomy, University of Calgary, Calgary, AB T2N 1N4, Canada
        \and Lahore University of Management Sciences (LUMS), Lahore, Pakistan
        \and NCBC at NED University of Engineering and Technology, Karachi, Pakistan
        \and University of Hull, Cottingham Road, Kingston-upon-Hull, HU6 7RX, UK
            }

   \date{Received September 30, 20XX}
 
  \abstract%
  {Sources detected by recent large-area radio surveys require robust classification and characterisation. The Evolutionary Map of the Universe (EMU) is a prime example of machine-learning (ML) techniques efficiently enriching existing catalogues with classifications and derived physical properties that would be impractical to obtain from spectroscopy or colour selection alone.}
  {We aim to build a labelled sample of sources in the EMU pilot survey (EMU-PS, at $\qty{944}{\MHz}$, with rms $\sim \qty{30}{\micro\jansky/\beam}$ over $\qty{270}{\deg\squared}$ in the southern sky) classified as being dominated by active galactic nuclei (AGNs) or non-AGN processes including star-formation. We will then use this sample to examine radio properties and redshift-dependent trends.}
  {We applied a pre-trained multi-stage ML pipeline to optical-to-infrared photometry of EMU-PS sources to derive classifications and photometric redshifts, which we used to compute additional radio properties, including luminosities.}
  {We increase the known number of AGNs and non-AGNs in the EMU-PS by more than a factor of $\num{15}$ (recall \qty{96}{\percent}) up to $z \sim \num{5}$, and by nearly a factor of $\num{100}$ (recall \qty{94}{\percent}) for $z \lesssim \num{1}$, respectively. In extrapolating these numbers for the full EMU survey, we expect to recover more than $\num{3.4}$ million AGN candidates. A non-negligible low-redshift subsample of selected non-AGNs shows radio luminosities in excess of those expected from star formation alone, revealing detachment between photometric and radio classification. Radio-derived properties show expected trends for luminosities and a lack of evolution for in-band ($\qtyrange[range-units=single, range-phrase=\mbox{--}]{800}{1088}{\MHz}$) spectral indices.}
  {The ML-based pipeline provides a scalable way to produce large catalogues for the ongoing EMU survey and future Square Kilometre Array observations and reveals that multi-wavelength analysis can uncover radio populations and trends often missed by colour cuts or spectroscopy alone.}


   \maketitle
   \nolinenumbers

\section{Introduction}\label{sec:intro}

    Astrophysics is entering a new era in the exploration of the high-redshift radio Universe. New surveys, including the Low Frequency Array \citep[LOFAR;][]{2013A&A...556A...2V} Two-metre Sky Survey \citep[LoTSS;][]{2017A&A...598A.104S} and the Evolutionary Map of the Universe \citep[EMU;][]{2011PASA...28..215N, 2025PASA...42...71H} are pushing the boundaries of radio active galactic nucleus (AGN) populations beyond unit redshift and, for radio star-forming galaxies (SFGs), to cosmologically relevant distances \citep{2019PASA...36...24L, 2024PASA...41...27G, 2026MNRAS.tmp..177W, 2026MNRAS.tmp..412W}.
    
    New radio surveys allow the unprecedented detection of high-redshift sources and probe SF rates up to $\qty{e-1}{\msun.\yr^{-1}}$ or $L_{\qty{144}{MHz}} \lesssim \qty{e21}{\W.\Hz^{-1}}$ at $z \sim \num{1}$ \citep{2021A&A...648A...6S}. These limits are expected to be surpassed by the future Square Kilometre Array Observatory \citep[SKAO;][]{2019arXiv191212699B}, improving our knowledge of both the far and faint radio Universe.

    Within this context, EMU is expected to reveal more than $\num{20}$ million extragalactic sources \citep{2025PASA...42...71H}, potentially revolutionising the statistical analysis of radio emission. EMU is being conducted with the Australian Square Kilometre Array Pathfinder \citep[ASKAP;][]{2021PASA...38....9H}, a precursor of the SKAO, to observe all of the southern sky (i.e.\ $\qty{2 \pi}{\steradian}$) up to a declination of $\ang[retain-explicit-plus]{+7;;}$ in some regions, with an rms of $\sim \qty{30}{\micro\jansky/\beam}$ in the $\qtyrange[range-units=single, range-phrase=\mbox{--}]{800}{1088}{\MHz}$ band centred at $\qty{944}{\MHz}$. While observations of the main EMU survey are in progress\footnote{At the time of writing, more than $\qty{50}{\percent}$ of observations are complete.}, an early pilot survey, mimicking its final conditions, was completed. The Pilot Survey of EMU \citep[EMU-PS;][]{2021PASA...38...46N} covered an area of $\qty{270}{\deg\squared}$ in the southern sky with a depth of $\qtyrange[range-units=single, range-phrase=\mbox{--}]{25}{30}{\micro\jansky/\beam}$ rms and a spatial resolution of $\sim \qty{18}{\arcsec}$\footnote{The native EMU-PS image resolution is $\sim \qty{11}{\arcsec}$.} within $\ang{301;;} < \mathrm{R.A.} < \ang{336;;}$ and $\ang{-63;;} < \mathrm{Dec.} < \ang{-48;;}$.

    One key goal is to identify sources harbouring AGNs, whose energetic output strongly affects their host galaxies and environments \citep[e.g.][]{2022MNRAS.513..256D}. Thus, it is critical to determine whether the emission is dominated by AGN or star formation (SF) activity. A number of methods have been developed using spectroscopy, photometry, and other indicators to identify AGN emission and constrain their properties.

    Using radio information and a redshift estimate, AGN can be selected based on their radio luminosity. If at a given redshift the luminosity of a source is higher than a certain limit, it can be classified as being dominated by AGN emission as SF activity alone is deemed not capable of reaching such levels \citep{2014MNRAS.442..682M, 2024A&A...685A..79W}. An additional selection method is based on radio morphology \citep{2023NewAR..9701685N}. Depending on how concentrated the emission is and on its shape, radio sources can be classified into several types and subclasses \citep{2021Galax...9...85R, 2025PASA...42..105B}. However, based on the source luminosity, its distance, and the characteristics of the observations, morphology is not a reliable observable to assess the nature of a radio detection \citep[e.g.][]{2015MNRAS.453.1644P, 2019MNRAS.488.2701M}.

    Complementing radio surveys, photometric colours place sources in different regions of the parameter space based on how red or blue they appear \citep[e.g. AGN, SF, or passive galaxies;][]{2010AJ....140.1868W, 2021A&A...649A..27R}. Widely used mid-infrared (MIR) criteria include colours from \emph{Spitzer} \citep{2004ApJS..154..166L, 2005ApJ...631..163S, 2012ApJ...748..142D}, the Wide-field Infrared Survey Explorer \citep[\emph{WISE};][]{2010AJ....140.1868W}, and the \emph{James Webb} Space Telescope \citep[JWST;][]{2012ApJ...754..120M, 2014A&A...562A.144M, 2017ApJ...849..111K, 2023ApJ...959L...7K}. Alternatively, multiple broad- or narrow-band colours can be used to select sources and estimate their photometric redshifts \citep[e.g.][]{2001A&A...365..660W, 2001A&A...365..681W, 2015MNRAS.448.2665W}. However, colour selection suffers from degeneracies and overlaps that place multiple source types in the same parameter-space region \citep[e.g.][]{2014A&A...562A.144M, 2023A&A...679A..76D}.

    For spectral-energy-distribution (SED) fitting tools, the inclusion of AGN components in the template bases helps to determine the fraction of radiation produced by their central engine. Besides source classification, these tools can provide further properties of AGNs and their hosts such as stellar masses or SF rates. Example SED fitting tools with AGN components include \texttt{SED3FIT} \citep{2013A&A...551A.100B}, \texttt{AGNfitter} \citep{2016ApJ...833...98C, 2024A&A...688A..46M}, \texttt{CIGALE} \citep{2019A&A...622A.103B, 2022ApJ...927..192Y}, \texttt{ProSpect} \citep{2020MNRAS.495..905R}, \texttt{Prospector} \citep{2021ApJS..254...22J}, and \texttt{SMART} \citep{2024MNRAS.531.2304V}. Despite their high reliability, which comes at the expense of long running times and the need for high-quality data, SED fitting methods can still reflect biases in their derived properties \citep{2022A&A...662A..86P, 2026ApJ...997..150F}.

    With spectroscopic information, sources can be separated upon the dominant emission lines' ionisation source, as quantified by line intensity ratios. For SF, emission is constrained by massive stars exciting their environments, while AGN produce harder X-ray photons ($\sim \unit{keV}$) independent of the host stellar content \citep{2006MNRAS.371..972S}. Commonly, BPT-VO diagrams \citep*[][]{1981PASP...93....5B, 1987ApJS...63..295V} are used, which combine the line ratios [\ion{O}{iii}]/H$\beta$, [\ion{N}{ii}]/H$\alpha$, [\ion{S}{ii}]/H$\alpha$, and [\ion{O}{i}]/H$\alpha$ to classify sources. Nevertheless, these methods are observationally expensive, with spectra available for only a fraction of sources.

    Active galactic nuclei can also be selected by combining measurements at separate wavelengths. One example is the connection between infrared (IR) and radio power. SF systems follow a tight radio--IR correlation \citep{1991ASPC...18..125H, 1992ARA&A..30..575C, 2001ApJ...554..803Y} and a radio excess relative to this relation implies additional AGN emission \citep[e.g.][]{1997MNRAS.289..824R, 2017MNRAS.468..217W}.

    Depending on the wavelength and the diagnostic used, a single source can appear dominated by AGNs or SF to different degrees. Obscuration and attenuation can suppress one component at some frequencies while leaving it visible at others. One particular example includes low- and high-excitation radio galaxies \citep[LERGs; HERGs;][]{1979MNRAS.188..111H, 1994ASPC...54..201L, 2007MNRAS.376.1849H, 2025MNRAS.543.4335A}. LERGs show strong radio emission but weak (or even absent) low-ionisation nuclear emission-line-region \citep[LINER;][]{1980A&A....87..152H} signatures in the optical and ultraviolet that is typical of passive galaxies. Opposite to them, HERGs display strong radio and optical emission but can also have SF features that come from their host galaxies. Thus, and despite differing in implementation complexity and overlapping in parameter space, the use of selection criteria from one single wavelength range can miss an important fraction of AGNs \citep{2019PASA...36...24L, 2024ApJ...971...68W}.

    As new surveys probe higher redshifts with deeper, wider observations, machine-learning (ML) techniques enable us to decipher information from large-area datasets, including the study of new source populations \citep[e.g.][]{2024ApJ...963..129M, 2024ApJ...964...39G}. In particular, ML methods have been used to select and characterise AGNs and their hosts, thus complementing traditional selection methods while also enabling physical insights to be drawn from the model outputs, including evolutionary studies. ML-based separation between AGNs and non-AGNs (nAGNs) has succeeded using photometry, spectra, or direct imaging \citep[e.g.][]{2024MNRAS.527.1163W, 2022MNRAS.515.6046P, 2025MNRAS.537.1028P}. Source characterisation has mainly focused on photometric redshifts \citep[e.g.][]{2022A&C....3900557L, 2023PASA...40...39L, 2025A&A...693A..73G}, although other properties have been estimated \citep[e.g. SF rates, stellar masses, and dust luminosities;][]{2020MNRAS.493.4808S, 2025A&A...699A.328A}. One key advantage of these methods is their use of information from a broad range of wavelengths, as SED fitting tools do, but within a fraction of the time other techniques take.

    As with photometric-colour methods, ML-based techniques present challenges based on the quality and representativeness of data. For photometric redshifts, training and inference data must adequately sample the relevant parameter space to avoid biases and poorly constrained predictions \citep{2023PASA...40....1H}. Missing or uncertain measurements can introduce further uncertainties into the predicted properties. In the case of AGNs, the observed SED may also be affected by the combined emission of the nucleus and host galaxy, obscuration, and intrinsic variability. In particular, measurements obtained at different epochs might not represent the same instantaneous SED, potentially increasing uncertainties in photometric redshift estimates \citep{2015A&A...574A.112D, 2025NCimR..48..537P}. These considerations motivate an assessment of how the available data affect predictions.

    In this work, we aim to expand the information contained in the EMU-PS source catalogue by applying a newly developed ML pipeline to determine the main radio emission contributor, i.e. the AGN or SF, and to estimate their redshifts. We used these classes and redshifts to analyse further properties, including radio luminosities and spectral indices. The pipeline and its models, described in \citet[][hereafter C23]{2023A&A...679A.101C} and Carvajal et al.\ (in prep.), use optical-to-MIR photometry from IR-detected sources in a given area as input to select radio-detectable AGNs and non-AGNs (rAGNs and rnAGNs, respectively).

    This paper is organised as follows. We briefly describe the prediction pipeline in Sect.~\ref{sec:section_02}. In Sect.~\ref{sec:section_03} we present the data used to obtain new sources. Section~\ref{sec:section_04} contains the results from the application of the prediction pipeline, and Sect.~\ref{sec:section_05} features their analysis. Finally, in Sect.~\ref{sec:section_06} we summarise our work and its conclusions. Throughout this work, we use the term non-AGN for any galaxy that does not present known properties of an AGN or that does not have AGN activity as its main emission source \citep[similar to a normal galaxy, as named by][]{1992ARA&A..30..575C}. We used AB magnitudes \citep{1974ApJS...27...21O} and adopted a flat $\Lambda$CDM cosmology, with $\Omega_{m} = \num{0.31}$, $\Omega_{\Lambda} = \num{0.69}$, and $H_{0} = \qty{67.7}{\km.\s^{-1}\mega\parsec^{-1}}$, as presented by the \citet{2020A&A...641A...6P}.

\section{Prediction pipeline}\label{sec:section_02}

    Models in the ML pipeline, fully described in \citetalias{2023A&A...679A.101C} (in particular, their Fig.~4) and Carvajal et al.\ (in prep.), are applied sequentially to the photometry of CatWISE2020-detected sources \citep[][hereafter CatWISE]{2021ApJS..253....8M}, since they were trained in mid-infrared-detected sources. The first step of the prediction pipeline probabilistically classifies CatWISE-detected (at $\qtylist[list-separator = {,\ \linebreak[0]}]{3.4; 4.6}{\micro\m}$) sources as AGNs and non-AGNs\@. Predicted AGNs are then passed to a second model that classifies them as radio-detectable or non-radio-detectable, and radio-detectable AGNs are sent to a third model that predicts their photometric redshifts. In parallel, sources initially classified as non-AGNs are passed to a fourth model that predicts their radio detectability, and radio-detectable non-AGNs are then processed by a fifth model that estimates their photometric redshifts. Additionally, the pipeline is modular, allowing individual stages to be omitted when the class or radio detectability of a source is already known.

    Each model uses ensemble learning, which takes the outputs of several base algorithms as input features, thus improving performance and reducing biases \citep{Dasarathy_ensemble_1979, WOLPERT1992241}. In particular, re-stacking \citep{michailidis2017investigating} is used, where the outputs of individual models are used alongside base features. Five tree-based \citep{breiman1984classification} and gradient-boosting \citep{FRIEDMAN2002367} models were selected per task. The following programs were used for classification: Random Forest \citep{2001MachL..45....5B}, Gradient Boosting Classifier \citep[GBC;][]{Friedman_2001}, Extra Trees \citep{Geurts2006}, Extreme Gradient Boosting \citep{Chen_2016}, and Category Boosting \citep[CatBoost;][]{NEURIPS2018_14491b75, Dorogush_2018}. For the redshift regression tasks, the models are the same but, instead of Gradient Boosting Classifier, Gradient Boosting Regressor \citep{Friedman_2001} was used. The best individually performing algorithm was selected for each task and used as a meta model with the output of the remaining ones as additional features. For the AGN--non-AGN classifier, CatBoost was used. For radio classification in AGNs, the meta model is GBC, while for the radio classification of non-AGNs, the selected model is Random Forest. For both redshift models, the meta model is Extra Trees.

    Machine-learning models generally require a sufficient number of input features to deliver reliable predictions \citep{PEARL01011978, Blumer_1987}. To reduce training costs, \citetalias{2023A&A...679A.101C} and Carvajal et al.\ (in prep.) selected only the most informative features for each model. Their initial feature set included $\num{12}$ optical-to-infrared magnitudes, $\num{66}$ colours derived from them, and a counter of detected magnitudes (Sect.~\ref{sec:section_03}). First, the authors computed the relative standard deviation \citep[RSD, a scaled measure of variance,][]{johnson1964statistics} for each feature, which allows the comparison of features with different ranges and values. Then, the Pearson correlation coefficient \citep{pearson1896correlation} was obtained for every pair of non-target features. For each highly correlated pair, the feature with the lower RSD (i.e.\ lowest variance) was removed. Applied independently to each model, this procedure reduced the number of features from $\num{79}$ to $\num{28}$ (full list in Sect.~\ref{sec:section_03}). To avoid discarding observations and harming the models and their predictions, missing photometry values were imputed using the limiting magnitude of each band, effectively working as upper limits for each of them (see Sect.~\ref{sec:section_03}).
    
    The first model outputs the predicted class (\texttt{0} for non-AGNs and \texttt{1} for AGNs) and the probability that the emission is AGN dominated, $p(\mathrm{AGN})$, from which the probability of star-formation dominance is $p(\mathrm{SF}) = 1 - p(\mathrm{AGN})$. The second-level models also return a class (\texttt{0} for radio undetectable and \texttt{1} for radio detectable) and the likelihood of radio detection, $p(\mathrm{radio \vert AGN})$ or $p(\mathrm{radio \vert SF})$, depending on the predicted class. Finally, the redshift models produce a single quantity, i.e. the photometric redshift, $z(\mathrm{radio,AGN})$ or $z(\mathrm{radio,SF})$, without an associated uncertainty because of the nature of the selected models, making these predictions more suitable for population-level than for object-by-object analyses (unless using the prediction intervals of Appendix~\ref{app:redshift_intervals}). In this work, focused on EMU-PS sources, we bypassed the radio detectability stage\footnote{For AGN selection, the threshold is $p(\mathrm{AGN}) = 0.349$. To select radio-detectable AGNs it is $p(\mathrm{radio \vert AGN}) = 0.205$, and to select radio-detectable non-AGNs it is $p(\mathrm{radio \vert SF}) = 0.200$. These values were chosen to maximise a measure of classification performance rewarding high completeness and low contamination in each class \citep{lipton_f1_2014}.}. Crucially, the models were trained on sources up to $z \sim \num{5}$ (Fig.~\ref{fig:HETDEX_DR1_true_z_hist}); hence, the resulting predictions are restricted to that range and should not be interpreted beyond it, which might affect downstream derived properties.

\section{Data}\label{sec:section_03}
    
    As we apply the existing pipeline models, the EMU-PS data must match the format used by \citetalias{2023A&A...679A.101C}. Those models were trained on CatWISE-detected sources with Pan-STARRS \citep{2016arXiv161205560C, 2020ApJS..251....7F} Data Release 1, Two Micron All Sky Survey \citep[2MASS;][]{2003tmc..book.....C, 2003yCat.2246....0C, 2006AJ....131.1163S}, and AllWISE \citep{2013wise.rept....1C} photometry. To account for differences in multi-wavelength coverage between the original training field and EMU-PS, we replaced Pan-STARRS photometry with Dark Energy Survey Data Release 2 \citep[DES-DR2;][]{2018PASP..130g4501M, 2021ApJS..255...20A} measurements of the southern hemisphere. Although their filter sets are not identical, \citet{2021ApJS..255...20A} validated their similarity, and \citet{2024MNRAS.533.2786J} showed that modest differences in sky coverage or depth do not noticeably degrade model performance.

    The EMU-PS value-added catalogue of compact sources\footnote{\url{https://doi.org/10.25919/exq5-t894}} \citep{2021PASA...38...46N} also offers counterparts and photometry, but we did not use them because the models were designed for sources initially detected by CatWISE\@. Instead, we constructed our own catalogue by selecting all $\num{10\,355\,457}$ CatWISE sources within the EMU-PS area and matching them independently within $\ang[angle-symbol-over-decimal]{;;1.5}$, $\ang[angle-symbol-over-decimal]{;;1.3}$, and $\ang[angle-symbol-over-decimal]{;;1.9}$ to DES-DR2, 2MASS, and AllWISE, respectively (see Appendix~\ref{app:search_radius}). The resulting catalogue contains photometry in the \textit{g}, \textit{r}, \textit{i}, \textit{z}, \textit{Y} (DES-DR2 Kron magnitudes, with central wavelengths at $\qtylist[list-separator = {,\ \linebreak[0]}]{473; 642; 784; 926; 1009}{\nm}$, respectively), \textit{J}, \textit{H}, \textit{Ks} (from 2MASS, at $\qtylist[list-separator = {,\ \linebreak[0]}, list-final-separator = {,\ \linebreak[0]\text{and }}]{1.235; 1.662; 2.159}{\micro\m}$), \textit{W1}, \textit{W2} (from CatWISE, centred at $\qtylist[list-separator = {,\ \linebreak[0]}]{3.4; 4.6}{\micro\m}$), and \textit{W3} and \textit{W4} bands (from AllWISE, at $\qtylist[list-separator = {,\ \linebreak[0]}]{12; 22}{\micro\m}$). Some sources lack measurements in one or more bands as they fall below detection limits, but we retained them and handled their missing values (Sect.~\ref{sec:section_03}).
    
    We then used a $\ang[angle-symbol-over-decimal]{;;1}$ radius \citep[following, for instance,][]{2015ApJS..221...12S} to match CatWISE sources to the Million Quasar\footnote{\url{http://quasars.org/milliquas.htm}} \citep[MILLIQUAS, v8;][]{2023OJAp....6E..49F} catalogue, the Quaia G20.5 \citep[the \textit{Gaia}-unWISE Quasar;][]{2024ApJ...964...69S} catalogue, and the southern-hemisphere spectroscopic redshift compilation \citep[][version \texttt{2024.06.20}\footnote{\url{https://zenodo.org/records/12728520}}, hereafter COMP]{2022A&C....3800510L, specz_compilation_zenodo}. Additionally, we used a $\ang[angle-symbol-over-decimal]{;;1.5}$ search radius to cross-match with DES-DR2 Photo-QSO \citep{2023ApJS..264....9Y} and the Visible and Infrared Survey Telescope for Astronomy \citep[VISTA;][]{2006Msngr.126...41E} EXtension to Auxiliary Surveys \citep[VEXAS;][]{2019A&A...630A.146S} Spec V2 \citep{2021A&A...651A..69K}. These catalogues provide the spectroscopic class used in the analysis (AGN or non-AGN).

    To avoid spurious classifications, we removed all confirmed stars from the EMU-PS catalogue. Together, the \emph{Gaia} measurements and the COMP catalogue contain the most comprehensive list of stars in the area. We discarded all $\num{18365}$ sources with \texttt{class = STAR} in COMP and, after cross-matching our catalogue with the \emph{Gaia} Data Release 3 \citep{2016A&A...595A...1G, 2023A&A...674A...1G} extragalactic content catalogue \citep{2023A&A...674A..41G} within $\ang[angle-symbol-over-decimal]{;;1}$, we discarded all $\num{2063601}$ sources with a high probability of being a star (\texttt{Pstar}, \texttt{PWD}, or \texttt{Pbin} over $\num{0.99}$). In total, $\num{2063611}$ stars were removed, leaving $\num{8291846}$ elements for analysis.

    \begin{table}[t]
        \setlength{\tabcolsep}{0.75pt}
        \centering
        \caption[Catalogue cross-matches in EMU pilot survey.]{Initial catalogue composition (non-stars detected by CatWISE, CW) and cross-match numbers.}\label{table:composition_catalogue_EMU}
        \begin{tabular}{c c S[table-format=8, detect-weight] S[table-format=8, detect-weight]}
        \toprule
        Step                & Survey        &  \multicolumn{2}{c}{Source number} \\
                            &               &  {CW}     & {CW+EMU-PS} \\
        \midrule
        \shortstack{Base\\sample} & CatWISE2020 & 8291846  & 129361    \\
        \cmidrule(r){2-4}
        \multirow{3}{*}{\shortstack{Photometry\\cross-match}} & DES-DR2 &  4791814 &   94528  \\
                            & AllWISE       &  2420195  &  78275    \\
                            & 2MASS         &    41584  &   8738    \\
        \cmidrule(r){2-4}
        \multirow{8}{*}{\shortstack{Source\\labelling}} & MILLIQUAS (AGN) &  454  &  215  \\
                            & Quaia (AGN)   &    12472  &   2007    \\
                            & DES-DR2 (AGN) &       46  &     26    \\
                            & COMP (AGN)    &     4766  &    694    \\
                            & (Total AGN)   &    12893  &   2103    \\
        \cmidrule(lr){3-4}
                            & VEXAS (non-AGN)   &     2039 &    876   \\
                            & COMP (non-AGN)    &     2929 &   1015   \\
                            & Sub-total non-AGN &     3097 &   1076   \\
                            & (AGN and non-AGN) &      151 &     88   \\
                            & (Total non-AGN)   &     2946 &   988    \\
        \cmidrule(r){2-4}
        \shortstack{Unlabelled\\sources} &  & 8276007  & 126270 \\ 
        \bottomrule
        \end{tabular}
    \end{table}

    Finally, following \citet{2021PASA...38...46N}, we used a $\qty{3}{\arcsec}$ search radius to cross-match CatWISE sources to the full EMU-PS compact sources catalogue, leading to an $\sim \qty{8}{\percent}$ false identification rate. The EMU-PS catalogue contains radio information at $\qty{944}{\MHz}$ for $\num{178\,921}$ compact sources above the $\qty{5}{\sigma}$ EMU detection limit, with a mean value of $\num{5} \times \qtyrange[range-units=bracket, range-phrase=\mbox{--}]{25}{30}{\micro\jansky} = \qtyrange[range-units=bracket, range-phrase=\mbox{--}]{125}{150}{\micro\jansky}$. Their integrated fluxes range roughly between $\qtylist[list-independent-prefix = true, exponent-to-prefix = true]{100;2e6}{\micro\jansky}$.\ Extended and complex sources \citep[$\sim \qty{19}{\percent}$ of sources in the EMU-PS;][]{2021PASA...38...46N} were not included, as a selection of optimal techniques for their extraction is in the development phase \citep{2025PASA...42..105B, 2026PASA...43...42B}. A summary of source number and counterparts in our catalogue is shown in Table~\ref{table:composition_catalogue_EMU}.

    Before applying the ML models to EMU-PS, we pre-processed its photometry. First, we targeted missing measurements with feature imputation as standard tree-based models cannot handle missing values \citep{1987samd.book.....L}. When a measurement is missing or too faint, we replaced it with the corresponding $\qty{5}{\sigma}$ limiting magnitude, mimicking the use of upper limits in SED fitting tools, following \citet{2021Galax...9...86C, 2023A&A...679A.101C}. This approach outperforms alternative imputation schemes \citep[e.g.][]{2022MNRAS.512.2099C, 2022MNRAS.514....1C} and removes the need for separate models trained on different band subsets. Unless stated otherwise, all subsequent calculations involving photometry used imputed values and must be treated accordingly, including the derived quantities.

    For DES-DR2 measurements, the $\qty{5}{\sigma}$ limits are $\qtylist[list-separator = {,\ \linebreak[0]}]{23.95; 23.65; 23.05; 22.35; 20.95}{\magnitude}$ in the \textit{g}, \textit{r}, \textit{i}, \textit{z}, and \textit{Y} bands, respectively. 2MASS limits are $\qtylist[list-separator = {,\ \linebreak[0]}]{17.45; 17.24; 16.59}{\magnitude}$ in \textit{J}, \textit{H}, and \textit{Ks}. CatWISE has limits of $\qtylist[list-separator = {,\ \linebreak[0]}]{20.13; 19.81}{\magnitude}$ for \textit{W1} and \textit{W2}, and AllWISE limits are $\qtylist[list-separator = {,\ \linebreak[0]}]{16.67; 14.62}{\magnitude}$ for \textit{W3} and \textit{W4}. Appendix~\ref{app:imputation_metrics} presents the number of measurements imputed in each non-radio band.

    From the original feature selection (Sect.~\ref{sec:section_02}), we computed $\num{25}$ colours after magnitude imputation: (\textit{g $-$ r}), (\textit{g $-$ i}), (\textit{g $-$ W1}), (\textit{g $-$ W2}), (\textit{r $-$ i}), (\textit{r $-$ z}), (\textit{r $-$ y}), (\textit{r $-$ J}), (\textit{i $-$ z}), (\textit{i $-$ y}), (\textit{z $-$ y}), (\textit{z $-$ W1}), (\textit{z $-$ W2}), (\textit{y $-$ J}), (\textit{y $-$ W1}), (\textit{y $-$ W2}), (\textit{J $-$ H}), (\textit{H $-$ K}), (\textit{H $-$ W3}), (\textit{Ks $-$ W3}), (\textit{Ks $-$ W4}), (\textit{W1 $-$ W2}), (\textit{W1 $-$ W3}), (\textit{W2 $-$ W3}), and (\textit{W3 $-$ W4}). These are complemented by two magnitudes, \textit{Ks} and \textit{W4}, and one additional feature, i.e. the number of non-radio bands in which a source is detected: \texttt{band\_num}. Values of \texttt{band\_num} range from $\num{2}$ (both CatWISE bands, the base catalogue) to $\num{12}$ (detections in all non-radio bands) and its usefulness in ML-based predictions has been tested previously \citep{zheng2018feature, 2021AJ....161..141S}. In this way, any dataset processed by the pipeline needs at least these $\num{28}$ features to obtain a prediction. These features were selected for the same reasons stated in Sect.~\ref{sec:section_02} regarding the minimum amount of information for useful predictions. 

    The final quantities needed to test the models are the targets (i.e.\ ground truth). The first being \texttt{class}, which can take the value $\mathtt{0}$ for non-AGNs, $\mathtt{1}$ for AGNs, and an invalid value (e.g. \texttt{NaN} or \texttt{-1}) for sources that are neither of the two (i.e.\ sources without a prior label). For radio detection, the target is \texttt{radio\_detect}, with $\mathtt{0}$ for no detection and $\mathtt{1}$ for radio detection (AGN or non-AGN)\footnote{All used sources are detected in EMU-PS; thus, \texttt{radio\_detect} is equal to \texttt{1} and not used in this work.}. For sources with previously determined redshifts, a measurement is provided through the feature \texttt{z\_comb}, which combines values from MILLIQUAS, Quaia, DES-DR2 Photo-QSO, and the COMP catalogues. Critically, we only assigned a source a \texttt{class} label (AGN or non-AGN) if it had an associated redshift, which allowed full comparison with all outputs of prediction pipeline. Additionally, when a source has been previously identified as both a non-AGN and an AGN, we favoured the AGN label.

\begin{figure}[t]
        \centering
        \includegraphics[width=0.96\columnwidth]{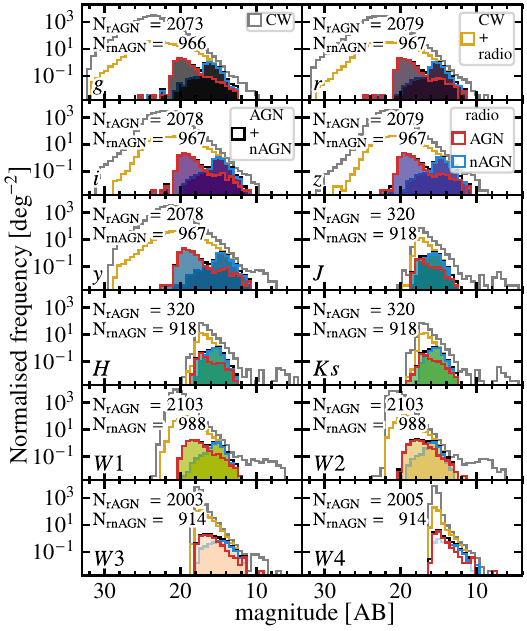}
        \caption{Magnitude distributions, normalised by area, of non-imputed, non-radio bands for radio AGNs (rAGNs, red), radio non-AGNs (rnAGNs, blue), all AGNs/non-AGNs (black), all EMU-PS-matched CatWISE sources (ochre), and all sources (grey) in the EMU-PS area. The upper left sections of each panel show rAGN and rnAGN counts, and the lower left corners show the selected magnitude.}\label{fig:hist_mags_rAGN_rSFG}
    \end{figure}

    Using DES-DR2, MILLIQUAS, VEXAS-DR2, Quaia, and COMP, we find that among the initial $\num{8291846}$ CatWISE-detected, non-star sources in the EMU-PS area, $\num{12\,893}$ are AGNs and $\num{2946}$ non-AGNs\@. Of these, $\num{129361}$ CatWISE sources have a counterpart in the EMU-PS catalogue, yielding $\num{2103}$ CatWISE radio-detected AGNs (rAGNs) and $\num{988}$ radio-detected non-AGNs (rnAGNs). In this work, we analysed the pipeline predictions for the CatWISE sources associated with an EMU-PS counterpart without applying any method to handle class imbalance. Figure~\ref{fig:hist_mags_rAGN_rSFG} shows the magnitude distribution for AGNs and non-AGNs\@. For further source analysis, we also compiled their in-band spectral indices, integrated $\qty{944}{\MHz}$ fluxes, and sizes of the deconvolved major axes from the fitted Gaussians in EMU-PS\@.

\begin{figure}[t]
        \centering
        \begin{overpic}[width=0.99\linewidth, percent]{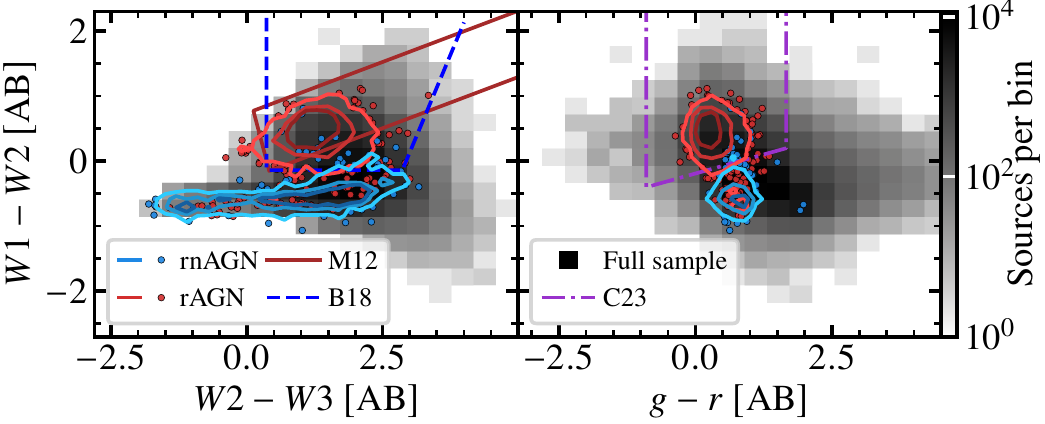}
            \put(11, 35){(a)\phantomsubcaption\label{fig:W2_W3_W1_W2_rAGN_rSFG}}
            \put(51, 35){(b)\phantomsubcaption\label{fig:W1_W2_g_r_rAGN_rSFG}}
        \end{overpic}
        \caption{Colour--colour diagrams of known rAGNs and radio non-AGNs (rnAGN) in EMU-PS\@. (a): WISE three-colour diagram.\ (b): (\textit{W1 $-$ W2}) versus (\textit{g $-$ r}) diagram. Blue, brown, and purple dashed lines: previous AGN-selection criteria by \citet{2018MNRAS.478.3056B}, \citet{2012MNRAS.426.3271M}, and \citetalias{2023A&A...679A.101C}, respectively. The greyscale density plot includes CatWISE-detected, non-imputed EMU-PS sources following the colour bar. Red (blue) contours contain non-imputed magnitudes of $\qtylist[list-separator = {,\ \linebreak[0]}]{50;75;95}{\percent}$ of rAGNs (non-AGNs) in the sample. Coloured points show outliers.}\label{fig:W2_W3_W1_W2_g_r_rAGN_rSFG}
    \end{figure}

    One way to assess the distribution of rAGNs and radio non-AGNs is through colour diagnostics. First, we used the colours (\textit{W1$-$W2}) and (\textit{W2$-$W3}), which are also known as the three-band WISE colour--colour diagram \citep[e.g.][]{2012MNRAS.426.3271M}. We present the location of our known sources (i.e.\ detected in EMU with a previous AGN or non-AGN label) in Fig.~\ref{fig:W2_W3_W1_W2_rAGN_rSFG}.

    The bulk of EMU-detected rAGNs and radio non-AGN lie in distinct areas of the three-band WISE colour--colour diagram, as also shown by \citet{2021MNRAS.506.3540P}. More importantly, their distributions match with previous AGN selection criteria. Known AGNs in EMU-PS are located mostly within the boundaries set by \citet[][hereafter M12]{2012MNRAS.426.3271M} and \citet[][hereafter B18]{2018MNRAS.478.3056B} and correspondingly, most known non-AGNs can be found outside such regions, matching the location of SFGs and passive or LINER galaxies \citep{2017ApJ...836..182J, 2023A&A...679A..76D}.

    We can also use the colour--colour criterion derived by \citetalias{2023A&A...679A.101C} from their analysis of ML-predicted sources. Figure~\ref{fig:W1_W2_g_r_rAGN_rSFG} shows the (\textit{g$-$r}) and (\textit{W1$-$W2}) distributions for radio AGNs and non-AGNs compared to their diagram. The bulk of the two populations are located in distinct regions of the parameter space.

\section{Model predictions}\label{sec:section_04}

    EMU-detected sources predicted to be AGN- or SF-dominated skip the radio-detectability step and are only subject to photometric redshift estimation. For classification assessment, we obtained and inspected their confusion matrices. For regression, we compared differences between predicted and true values from MILLIQUAS, Quaia, DES-DR2 Photo-QSO, and COMP\@.

    \subsection{Prediction assessment metrics}\label{sec:pred_metrics}

    The performance of classification tasks is evaluated by comparing new classes with true labels. The most basic tool is the confusion matrix \citep{pearson1904theory, Miller_1955}. It contains the number of correctly and incorrectly classified elements. Each sub-cell counts true positives (TPs), false positives (FPs), true negatives (TNs), and false negatives (FNs). These four quantities can be combined to obtain further classification metrics.

    Two widely used classification metrics are precision and recall. Precision \citep[also called purity or pertinence, $\mathcal{P}$;][]{Kent_1955} is the ratio of correctly classified positive elements over the total number of positively predicted sources. It is defined as
    \begin{equation}\label{eq:precision}
    \text{Precision} = \mathcal{P} = \dfrac{\text{TP}}{\text{TP} + \text{FP}} \,.
    \end{equation}
    \noindent Additionally, recall \citep[also known as completeness or true positive rate; $\mathcal{R}$,][]{Yerushalmy_1947} is the ratio of correctly recovered elements over the total number of recoverable objects. From the confusion matrix, this is
    \begin{equation}\label{eq:recall}
    \text{Recall} = \mathcal{R} = \dfrac{\text{TP}}{\text{TP} + \text{FN}} \,.
    \end{equation}

    As both predicted classes (AGNs and non-AGNs) are equally relevant, we combined each pair of metrics into a global precision and a recall category ($\mathcal{P}^{c}$ and $\mathcal{R}^{c}$, respectively). Since these metrics are defined as ratios, we used a weighted harmonic mean to join them using the inverse of the fraction of true samples as weights. Thus, the general formula for $\mathcal{P}^{c}$ or $\mathcal{R}^{c}$ can be written as
    \begin{equation}\label{eq:harmonic_mean}
        x^{c} = \left( \sum\limits_{i\in \{\mathrm{AGN}, \mathrm{non-AGN}\}} w_i x_i^{-1} \middle/ \sum\limits_{i\in \{\mathrm{AGN}, \mathrm{non-AGN}\}} w_i \right)^{-1},
    \end{equation}
    \noindent where $x_{i}$ is the selected metric and $w_{i}$, the weight, is the inverse of the prevalence, the fraction of known AGNs or non-AGNs over the total number of known sources in the sample, or
    \begin{equation}\label{eq:harmonic_weight}
        w_i = \dfrac{\mathrm{TN}_i + \mathrm{FP}_i + \mathrm{FN}_i + \mathrm{TP}_i}{\mathrm{TP}_i + \mathrm{FN}_i}.
    \end{equation}

    After an assessment of classification performance, we moved to redshift estimation, for which we adopted the normalised median absolute deviation (NMAD) and the outlier fraction ($\eta$). The NMAD \citep[$\sigma_{\text{NMAD}}$;][]{hoaglin1983understanding} relates true and predicted redshifts with a normalisation based on the non-linear nature of the redshift \citep{2000ApJ...538...29C, 2004A&A...421..913W}. It is defined as
    \begin{equation}\label{eq:nmad}
    \sigma_{\text{NMAD}} = \num{1.48} \times \text{median} \left| \dfrac{\Delta z}{1 + z_{\text{true}}} \right| \,,
    \end{equation}
    \noindent where the factor $\num{1.48}$ makes the metric comparable with a Gaussian standard deviation \citep[e.g.][]{2006A&A...457..841I} and $\Delta z = z_{\text{pred}} - z_{\text{true}}$ is the redshift dispersion.

    The outlier fraction is the ratio of sources that have a normalised redshift difference above a specific value \citep{2004A&A...421..913W}. We followed the limit first set by \citet{2006A&A...457..841I} of $\num{0.15}$. Its definition can be written as
    \begin{equation}\label{eq:outlier_fraction}
    \eta = \dfrac{1}{N} \times \mbox{\large$\#$} \left( \left| \dfrac{\Delta z}{1 + z_{\text{true}}} \right|  > 0.15 \right) \,,
    \end{equation}
    \noindent in which $\#$ indicates the number of sources fulfilling the condition in parentheses and $N$ is the total number of sources to compare. Both metrics can either be obtained independently for each class or combined for the complete set of predicted sources.  Further assessment metrics and their values for our predictions are described in Appendix~\ref{app:more_prediction_metrics}.

    \subsection{Selected sources}\label{sec:selected_sources}

\begin{figure}[t]
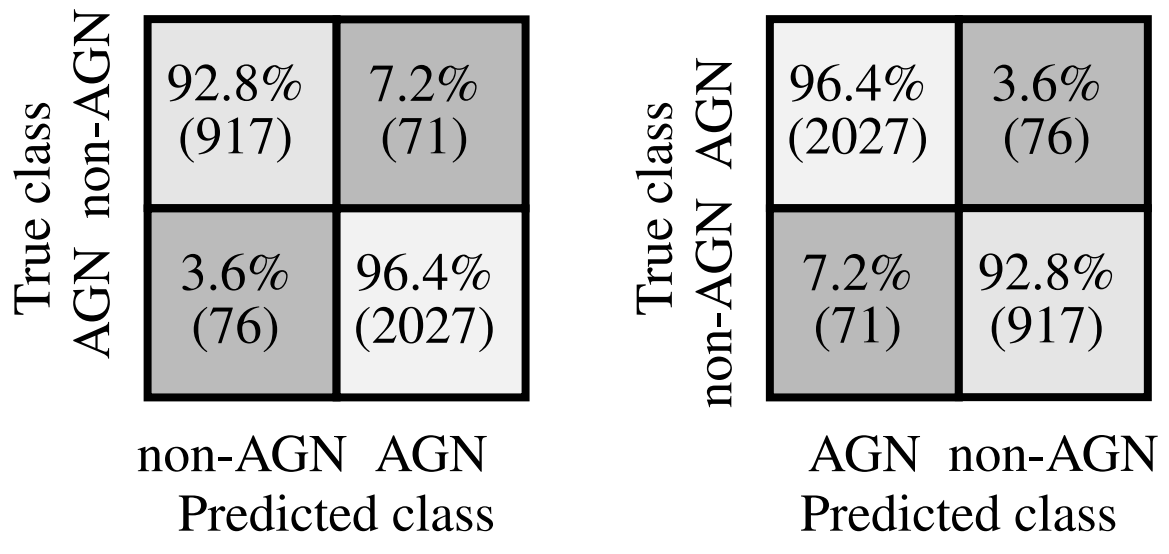

        \captionsetup{skip=0pt, belowskip=-6pt}
        \begin{center}
        \begin{subfigure}[t]{0.45\columnwidth}
        {
        \centering
        \confusionmatrixperc[true]{917}{71}{76}{2027}{AGN}{non-AGN}
        \caption{}\label{fig:conf_matrix_AGN}
        }
        \end{subfigure}
        \begin{subfigure}[t]{0.45\columnwidth}
        {
        \centering
        \confusionmatrixperc[true]{2027}{76}{71}{917}{non-AGN}{AGN}
        \caption{}\label{fig:conf_matrix_SFG}
        }
        \end{subfigure}
        \end{center}
        \caption{Confusion matrices for (a) AGN and (b) non-AGN source classifications. Columns represent predicted classes and rows true categories. Each cell shows a row-based percentage followed by its number of sources; these are also coded by cell colour.}\label{fig:AGN_conf_matrices}
    \end{figure}

    We applied the pipeline to the $\num{129361}$ CatWISE-detected sources with an EMU-PS counterpart. From them, $\num{30866}$ are predicted to be AGNs and $\num{2027}$ of those are confirmed AGN, while $\num{71}$ are confirmed non-AGNs\@. Accordingly, $\num{98495}$ sources are predicted to be non-AGNs, with $\num{917}$ of them confirmed non-AGNs and $\num{76}$ confirmed AGNs\@. For sources with previous classifications, Figs.~\ref{fig:conf_matrix_AGN} and~\ref{fig:conf_matrix_SFG} present their confusion matrices. The non-AGN confusion matrix is a rotated version of that for AGNs\@. Their values can be translated, for AGNs, into a recall of $\mathcal{R}^{\mathrm{AGN}} = \qty{96.4}{\percent}$ and a precision of $\mathcal{P}^{\mathrm{AGN}} = \qty{96.6}{\percent}$. For non-AGNs, their recall is $\mathcal{R}^{\mathrm{non-AGN}} = \qty{92.8}{\percent}$ and precision is $\mathcal{P}^{\mathrm{non-AGN}} = \qty{92.4}{\percent}$. The integrated classification has a combined precision of $\mathcal{P}^{c} = \qty{93.7}{\percent}$ and a combined recall of $\mathcal{R}^{c} = \qty{93.9}{\percent}$.

\begin{figure}[t]
        \centering
        \begin{overpic}[width=0.99\linewidth, percent]{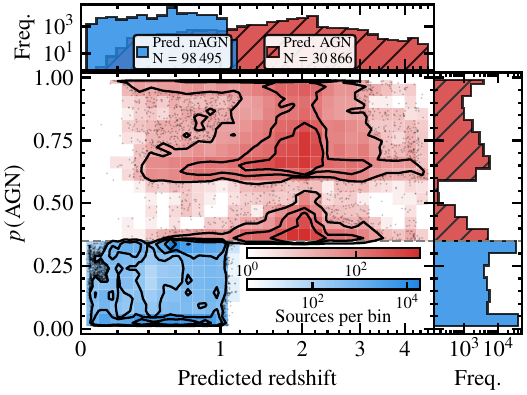}
            \put(76, 70){(a)\phantomsubcaption\label{fig:EMU_PS_all_pred_z}}
            \put(76, 56){(b)\phantomsubcaption\label{fig:pred_z_vs_class_prob}}
            \put(94, 56){(c)\phantomsubcaption\label{fig:EMU_PS_all_class_prob}}
        \end{overpic}
        \caption{Distribution of predicted (a) redshifts and (c) class probabilities for predicted non-AGNs (nAGNs, blue) and predicted AGNs (hatched red) in the sample. Legend shows the source number in each subset\@. (b) Two-dimensional distribution of photometric redshift versus class probabilities. Contours contain $\qtylist[list-separator = {,\ \linebreak[0]}]{50; 75; 95}{\percent}$ of the sources. Black points are outliers. Background bins show the source number following the colour bars. The dashed grey line shows the \citetalias{2023A&A...679A.101C} AGN-selection-probability threshold, $\num{0.349}$.}\label{fig:large_pred_z_vs_class_prob}
    \end{figure}

    For reference, we compared our results with those of \citet{2023A&A...675A.159K}, whose authors used a tree-based ML model to separate AGNs from non-AGNs (specifically, SFGs) using ultraviolet-to-radio photometry from more than $\num{60000}$ sources in the LoTSS Deep Fields \citep{2021A&A...648A...1T, 2021A&A...648A...2S}. From predictions over close to $\num{2500}$ sources, they report $\mathcal{R}^{\mathrm{AGN}} = \qty{79}{\percent}$, $\mathcal{P}^{\mathrm{AGN}} = \qty{87}{\percent}$, $\mathcal{R}^{\mathrm{non-AGN}} = \qty{96}{\percent}$, and $\mathcal{P}^{\mathrm{non-AGN}} = \qty{92}{\percent}$. Despite using a narrower wavelength range and shallower measurements in some bands, our performance is comparable to and even exceeds theirs for AGN selection, which might be a consequence of our use of models trained in a larger area. The distribution of class probabilities and photometric redshifts of the full sample, summarising results of the prediction pipeline, is presented in Fig~\ref{fig:large_pred_z_vs_class_prob}.

    Most predicted non-AGN photometric redshifts (Fig.~\ref{fig:EMU_PS_all_pred_z}) are uniformly distributed in the $\num{0} \la z \la \num{1}$ range, peaking at $z \approx \num{0.58}$. This distribution matches expectations for radio sources with no strong cosmological evolution \citep{2016MNRAS.460....2P, 2020MNRAS.496.1706S}, confirming minimal selection effects in the predicted non-AGN population. In contrast, AGN redshifts are in a wider range ($\num{0} \la z \la \num{5}$) with a noticeable peak at $z \approx \num{2.2}$ and a drop just above $z \sim \num{0}$. This peak coincides with the height of co-moving AGN luminosity activity density at $z \sim \num{2}$ \citep{2014ARA&A..52..589H, 2015ApJ...813L...8M, 2022ApJ...924..133R}. As Fig.~\ref{fig:EMU_PS_all_class_prob} shows, SF probabilities, $p(\mathrm{SF})$, have two peaks, one close to $p(\mathrm{SF}) = \num{1.0}$ and another close to the pipeline threshold of $p(\mathrm{SF}) = \num{1.0} - p(\mathrm{AGN}) = \num{1.0} - \num{0.349} = \num{0.651}$. On the other hand, AGN class probabilities split into two sub-populations separated by $p(\mathrm{AGN}) \sim \num{0.55}$, pointing to a fraction of low-confidence predictions. In Appendix~\ref{app:new_agn_threshold}, we explore some effects of the re-classification of these low-confidence AGNs as non-AGNs\@.

\begin{figure}[t]
        \centering
        \begin{overpic}[width=0.99\linewidth, percent]{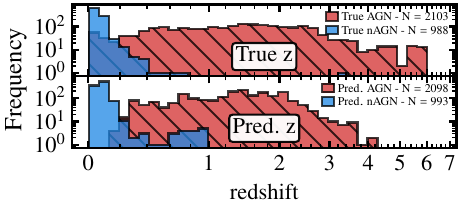}
            \put(93, 31){(a)\phantomsubcaption\label{fig:EMU_PS_true_z}}
            \put(93, 15){(b)\phantomsubcaption\label{fig:EMU_PS_pred_z}}
        \end{overpic}
        \caption{Redshift distribution for originally confirmed sources in EMU-PS\@. (a): True redshifts for true non-AGNs (nAGNs, blue) and true AGNs (hatched red).\ (b): ML-predicted photometric redshifts for predicted non-AGNs (blue) and AGNs (hatched red). Legends show the number of sources in each subset.}\label{fig:EMU_PS_true_pred_z}
    \end{figure}

    We restricted ourselves to sources with an existing classification, which we call true sources (as described in Sect.~\ref{sec:section_03}). The distributions of predicted and previously available redshifts for true sources are shown in Fig.~\ref{fig:EMU_PS_true_pred_z}, which have similar shapes and peak at the same values: AGNs at $z \sim \num{2}$ and non-AGNs at $z \sim \num{0.25}$. Additionally, non-AGNs in Fig.~\ref{fig:EMU_PS_pred_z} show a secondary peak at $z \lesssim \num{1}$, which is mostly populated by misclassified AGNs that are pushed to the highest redshift possible by the non-AGN redshift model.

\begin{figure}[t]
        \centering
        \begin{subfigure}[t]{0.48\columnwidth}%
            \centering
            \begin{overpic}[width=0.99\linewidth, percent]{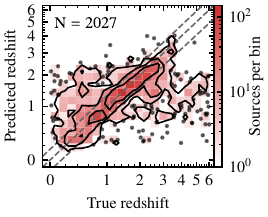}
                \put(-2, 80){(a)\phantomsubcaption\label{fig:predicted_z_rAGN_true_rAGN}}
            \end{overpic}
        \end{subfigure}%
        \begin{subfigure}[t]{0.50\columnwidth}%
            \centering
            \begin{overpic}[width=0.99\linewidth, percent]{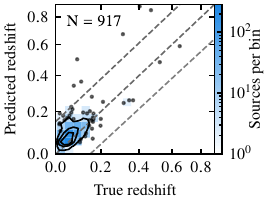}
                \put(0, 74){(b)\phantomsubcaption\label{fig:predicted_z_rSFG_true_rSFG}}
            \end{overpic}
        \end{subfigure}
        \caption{True versus predicted redshifts for correctly predicted (TP) (a) AGNs and (b) non-AGNs\@. Contours contain $\qtylist[list-separator = {,\ \linebreak[0]}]{50;75;95}{\percent}$ of the sources (total number in upper left corner). Background bins show the source number following colour bars. Dashed lines show $x = y$ line and outlier limits (Eq.~\ref{eq:outlier_fraction}), with outliers in black.}\label{fig:predicted_z_true_sources_EMU_PS}
    \end{figure}

    Finally, Fig.~\ref{fig:predicted_z_true_sources_EMU_PS} shows the predicted versus true redshifts for correctly predicted sources (i.e.\ sources in the TP quadrant of Figs.~\ref{fig:conf_matrix_AGN} and~\ref{fig:conf_matrix_SFG}). The majority of both correctly predicted AGNs and non-AGNs are located around the identity line and more than $\qty{95}{\percent}$ of them lie within the outlier limits.

    For the $\num{2027}$ and $\num{917}$ correctly predicted AGNs and non-AGNs in Fig.~\ref{fig:predicted_z_true_sources_EMU_PS}, we obtained prediction metrics. The prediction pipeline was trained with a low fraction of high-redshift AGNs ($z \gtrsim \num{4.5}$, see Fig.~\ref{fig:HETDEX_DR1_true_z_hist}). Thus, all its predictions are biased against such values, impacting the metrics. For radio AGNs predicted to be AGNs, the NMAD is $\sigma_{\text{NMAD}}^{\mathrm{AGN}} = \num{0.103}$ and the outlier fraction is $\eta^{\mathrm{AGN}} = \qty{31.0}{\percent}$. For non-AGNs predicted as non-AGNs, these values are $\sigma_{\text{NMAD}}^{\mathrm{non-AGN}} = \num{0.030}$ and $\eta^{\mathrm{non-AGN}} = \qty{1.0}{\percent}$\footnote{For the $\num{76}$ misclassified AGNs, the metrics are $\sigma_{\text{NMAD}}^{\mathrm{AGN}} = \num{0.043}$ and $\eta^{\mathrm{AGN}} = \qty{17.1}{\percent}$, while for the $\num{71}$ non-AGNs classified as AGNs, $\sigma_{\text{NMAD}}^{\mathrm{AGN}} = \num{0.657}$ and $\eta^{\mathrm{AGN}} = \qty{98.6}{\percent}$.}. The combined redshift prediction for EMU-detected AGNs and non-AGNs gives $\sigma_{\text{NMAD}}^{c} = \num{0.066}$ and $\eta^{c} = \qty{21.6}{\percent}$.

    We contrasted our redshift prediction metrics with those of \citet{2025ApJS..277...41K}, whose authors trained a tree-based model with optical-to-infrared photometry from sources (including AGNs and non-AGNs) in the North-Ecliptic-Pole-Wide (NEPW) field, achieving $\sigma_{\text{NMAD}}^{c} = \num{0.028}$ and $\eta^{c} = \qty{7.3}{\percent}$. They split the outlier fraction by source, with $\eta^{\text{AGN}} = \qty{40}{\percent}$ and $\eta^{\text{non-AGN}} = \qty{9}{\percent}$. Their strong non-AGN performance drove these results, whereas ours are superior for AGNs\@. Similarly, \citet{2021A&A...648A...4D} combined optical-to-infrared template fitting and ML techniques to estimate a photometric redshift of close to $\num{5}$ million sources in the LoTSS Deep Fields, resulting in $\sigma_{\text{NMAD}}^{\text{non-AGN}} = \num{0.019}$, $\sigma_{\text{NMAD}}^{\text{AGN}} = \num{0.058}$, $\eta^{\text{non-AGN}} = \qty{2}{\percent}$, and $\eta^{\text{AGN}} = \qty{19}{\percent}$. These values were driven by their large training set and aided by the addition of $u$-band photometry and photometric uncertainties in their ML modelling.

\section{Analysis of predictions}\label{sec:section_05}

    In this section we show some properties of the predicted sources, as well as a discussion on their values.

    \subsection{Colour analysis}\label{sec:colour_analysis}

\begin{figure}[t]
        \centering
        \begin{subfigure}[t]{0.68\columnwidth}
            \centering
            \begin{overpic}[width=0.99\linewidth, percent]{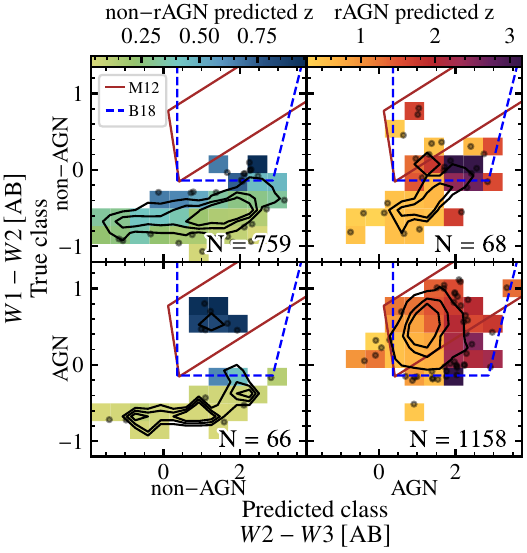}
                \put(01, 95){(a)\phantomsubcaption\label{fig:WISE_CM_EMU_sources}}
            \end{overpic}
        \end{subfigure}\\[1ex]%
        \begin{subfigure}[t]{0.68\columnwidth}
            \centering
            \begin{overpic}[width=0.99\linewidth, percent]{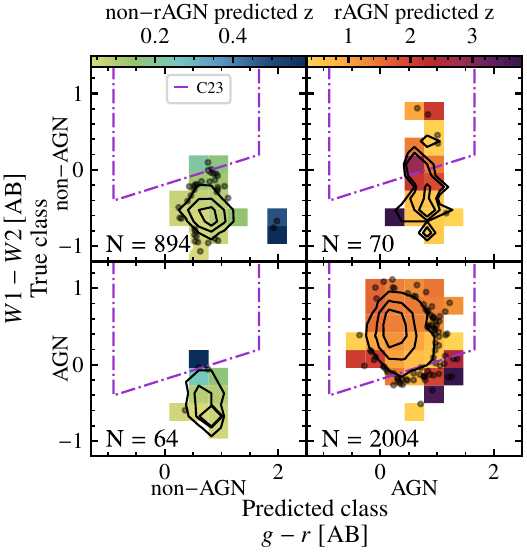}
                \put(01, 95){(b)\phantomsubcaption\label{fig:C23_CM_EMU_sources}}
            \end{overpic}
        \end{subfigure}
        \caption{Non-imputed (a) WISE three-band and (b) (\textit{W1$-$W2}) versus (\textit{g$-$r}) colour--colour diagrams for known sources in EMU-PS\@. Each sub-cell represents elements of the confusion matrix (Fig.~\ref{fig:conf_matrix_AGN}) for AGN-selection prediction (plotted number in the lower corner of each quadrant). Binned plots show median predicted $z$ following colour bars. Contours limit $\qtylist[list-separator = {,\ \linebreak[0]}]{50;75;95}{\percent}$ of the sources in each quadrant, and black points are outliers. Previous AGN-selection criteria from \citetalias{2012MNRAS.426.3271M}, \citetalias{2018MNRAS.478.3056B}, and \citetalias{2023A&A...679A.101C}, included for reference.}\label{fig:CM_EMU_sources}
    \end{figure}

    In Fig.~\ref{fig:CM_EMU_sources} we present the three-band WISE and (\textit{W1$-$W2}) versus (\textit{g$-$r}) colour--colour diagrams for previously classified sources, which are split over their matching confusion matrices (see Figs.~\ref{fig:conf_matrix_AGN} and~\ref{fig:conf_matrix_SFG}).

    Most correctly predicted (TP) AGNs in Fig.~\ref{fig:CM_EMU_sources} populate the \citetalias{2023A&A...679A.101C} and \citetalias{2018MNRAS.478.3056B} regions with a smaller fraction of sources covered by the \citetalias{2012MNRAS.426.3271M} criterion.\ \citetalias{2012MNRAS.426.3271M} analysed WISE colours of X-ray-selected AGNs out to $z \sim \num{2}$ with mean redshifts in the $z = \numrange{0.3}{0.7}$ range. Conversely, \citetalias{2018MNRAS.478.3056B} studied IR properties of AGNs taking part in galaxy mergers and dual AGN hosts at $z \lesssim \num{1}$. Therefore, it is expected that the \citetalias{2012MNRAS.426.3271M} and \citetalias{2018MNRAS.478.3056B} criteria cannot cover all predicted sources in our sample as they only study (independently) unobscured X-ray and dusty IR galaxies. Regardless, they still show a good coverage of our predicted populations.
    
    Additionally, correctly predicted AGNs in the WISE three-band plot of Fig.~\ref{fig:WISE_CM_EMU_sources} show a mild gradient of median photometric redshift that might hint at a colour evolution. More specifically, the progression of predicted redshift follows a horseshoe shape, starting close to the ($\num{1.0}$, $\num{0.0}$) point, going up to ($\num{1.5}$, $\num{1.0}$), and returning to ($\num{2.0}$, $\num{-0.2}$). This evolution resembles the estimated WISE colours of AGNs at $z = \numrange[range-phrase = \mbox{--}]{0}{2}$ presented by \citetalias{2012MNRAS.426.3271M} and their Fig.~5; the authors analysed galaxy templates from \citet{2008ApJ...675..960P}. When the AGN fraction is over $\qty{80}{\percent}$ in their templates, WISE colours also present a horseshoe-like behaviour. This parallelism suggests that, in EMU-PS, AGNs predicted by the pipeline also have low SF contamination.

    When analysing correctly predicted non-AGNs (i.e. TN, top left quadrant of Figs.~\ref{fig:WISE_CM_EMU_sources} and~\ref{fig:C23_CM_EMU_sources}), their distribution changes accordingly. Most sources lie well outside typical AGN regions. In Fig.~\ref{fig:WISE_CM_EMU_sources}, correctly labelled non-AGNs still show a gradient in median predicted redshift, although it is weaker than for predicted AGNs\@. Their photometric redshift increases with (\textit{W1}$-$\textit{W2}) at fixed (\textit{W2$-$W3}), which is consistent with the simulations of low-AGN-fraction sources by \citetalias{2012MNRAS.426.3271M} and ultra-luminous IR galaxies (ULIRGs) and starburst galaxies from \citet{2021A&A...649A..27R}, which was based on several MIR selection criteria over galaxy SED templates from \citet{2007ApJ...663...81P}. A further analysis of selected colour evolution with predicted redshift is presented in Appendix~\ref{app:colour_evol}.
    
\begin{figure}[t]
        \centering
        \includegraphics[width=0.71\linewidth]{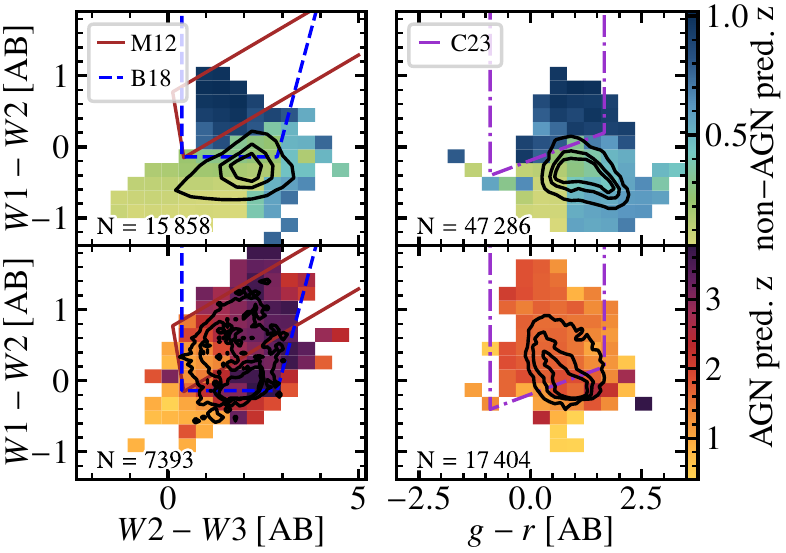}
        \caption{Non-imputed WISE three-band diagram (left column) and (\textit{W1$-$W2}) versus (\textit{g$-$r}) colour--colour diagram (right column) for predicted non-AGNs (upper row) and AGNs (lower row) in EMU-PS, including previously labelled sources. Binned plots show median predicted $z$ for sources following colour bars. Contours contain $\qtylist[list-separator = {,\ \linebreak[0]}]{50; 75; 95}{\percent}$ of the sources in each quadrant (total number in the lower left corner). AGN-selection criteria from \citetalias{2012MNRAS.426.3271M} (blue), \citetalias{2018MNRAS.478.3056B} (brown), and \citetalias{2023A&A...679A.101C} (purple) included for reference. Outliers were omitted for clarity.}\label{fig:WISE_C23_CM_EMU_PS_pred_sources}
    \end{figure}

    In the case of incorrectly predicted sources (i.e.\ elements in the FP and FN, also called off-diagonal, quadrants), they lie both inside and outside the AGN regions. Thus, there is a mix of sources with WISE and DES-DR2 AGN and non-AGN properties, which is expected for sources that confused the models in their prediction. Another point to highlight is that sources in the AGN regions of both FN quadrants of Fig.~\ref{fig:CM_EMU_sources} have the highest predicted photometric redshifts. Notably, the diagrams in Fig.~\ref{fig:CM_EMU_sources} only account for two colours at a time, while the prediction pipeline uses more magnitudes and colours, extracting richer information from the analysed sources. Thus, it is possible to have sources in places not expected in colour--colour diagrams while also achieving a correct prediction from the pipeline. Finally, all predicted sources in our sample (see Fig.~\ref{fig:large_pred_z_vs_class_prob}) can be placed in the WISE and (\textit{W1$-$W2}) versus (\textit{g$-$r}) colour--colour diagrams to study their distribution, as shown in Fig.~\ref{fig:WISE_C23_CM_EMU_PS_pred_sources}.

    The left columns of Fig.~\ref{fig:WISE_C23_CM_EMU_PS_pred_sources} are in line with results in Fig.~\ref{fig:WISE_CM_EMU_sources}. Most predicted AGNs are in the region defined by \citetalias{2018MNRAS.478.3056B}, while the majority of predicted non-AGNs are below the AGN wedge by \citetalias{2012MNRAS.426.3271M}. Additionally, mean AGN predicted redshifts show a horseshoe-like evolution as those in Fig.~\ref{fig:CM_EMU_sources}, although not as strongly as correctly predicted sources. The case of predicted non-AGNs is not as clear as with AGNs, where part of their distribution peak bleeds into the region by \citetalias{2018MNRAS.478.3056B} and no clear median photometric redshift evolution is seen although the highest redshift sources are well within the AGN region. In the right-hand side panels of Fig.~\ref{fig:WISE_C23_CM_EMU_PS_pred_sources}, predicted sources follow a similar pattern but with the \citetalias{2023A&A...679A.101C} region. However, in this case, a non-negligible fraction of predicted AGNs lie outside the AGN zone, and the non-AGNs with the highest redshifts are located inside the AGN region.

\begin{figure}[t]
        \centering
        \includegraphics[width=0.71\linewidth]{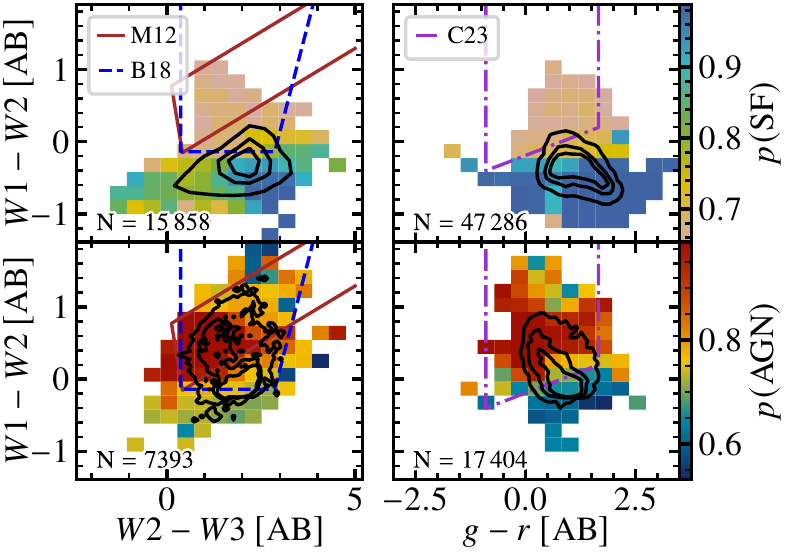}
        \caption{Same as Fig.~\ref{fig:WISE_C23_CM_EMU_PS_pred_sources} but coloured by median class probability, $p(\mathrm{SF})$, for the top row and $p(\mathrm{AGN})$ for the bottom panels.}\label{fig:WISE_C23_CM_EMU_PS_pred_sources_prob}
    \end{figure}

    From Fig.~\ref{fig:WISE_C23_CM_EMU_PS_pred_sources}, it is clear that the colour does not behave as expected. We tested the pertinence of our analysis by studying the colour evolution as a function of class probability: $p(\mathrm{SF})$ or $p(\mathrm{AGN})$. In Fig.~\ref{fig:WISE_C23_CM_EMU_PS_pred_sources_prob} we include the median class probability of predicted sources. It can be seen, for example, that predicted non-AGNs located within AGN boundaries have the lowest median probabilities among all candidates, i.e. close to $p(\mathrm{SF}) = \num{0.6}$. Thus, the prediction model is not certain of its classification for these sources. A similar effect is seen for predicted AGNs outside their expected loci. Their median probabilities show a low degree of certainty, although less uniformly than non-AGNs\@.

    To reiterate, the location of sources in these diagrams (Figs.~\ref{fig:WISE_C23_CM_EMU_PS_pred_sources} and~\ref{fig:WISE_C23_CM_EMU_PS_pred_sources_prob}) can be taken as a low-complexity diagnostic tool that is prone to uncertainties. Models in the prediction pipeline use information from several wavelengths and surveys, surpassing the capabilities of individual criteria, such as those analysed in this section. Thus, the use of multi-wavelength, multi-survey tools, such as ML-based classifiers, provides a broader coverage of the properties of sources. Also, sources with low-class probabilities are backed by enough information to include them in further analyses. Moreover, the pipeline used in this work, with its multi-wavelength nature, can be applied with different classification thresholds for AGNs and non-AGNs\@. Thus, and depending on the research goals, more stringent limits can be set.

    In addition, the samples used by \citetalias{2012MNRAS.426.3271M} and \citetalias{2018MNRAS.478.3056B} consider AGNs at lower redshifts than our predicted populations. As shown in Fig.~\ref{fig:WISE_C23_CM_EMU_PS_pred_sources}, predicted AGNs outside their expected regions tend to have higher predicted redshifts and might not need to follow the AGN criteria. These results are backed by \citet{2023A&A...679A..76D}, whose authors consider the bulk of their AGN sample to lie outside said MIR criteria, as they might be biased towards obscured and luminous sources.

    \subsection{EMU-PS quasar candidates}

\begin{figure}[t]
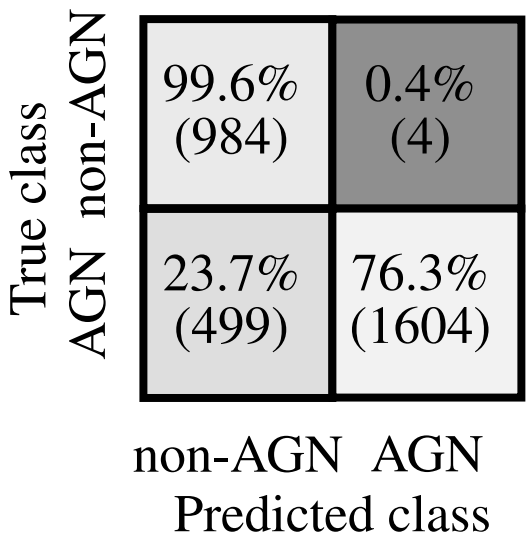

        \centering
        \confusionmatrixperc[true]{984}{4}{499}{1604}{AGN}{non-AGN}
        \caption{Confusion matrix for classification between AGNs (QCs) and non-AGNs (non-QCs) as provided by the EMU-PS catalogue of compact sources. The description is as in Fig.~\ref{fig:AGN_conf_matrices}.}\label{fig:QC_EMU_PS_conf_matrix}
    \end{figure}

    The EMU-PS catalogue of compact sources includes a flag for quasar candidates (QCs). These were produced by cross-matching DES-detected \citep[DES-DR1;][]{2018ApJS..239...18A} EMU-PS sources with the \emph{Gaia} Early Data Release 3 \citep[Gaia EDR3;][]{2021A&A...649A...1G} using a search radius of $\qty{0.2}{\arcsec}$. These sources were also cross-matched to CatWISE and AllWISE catalogues with a $\qty{1}{\arcsec}$ search radius and then placed in the WISE three-colour diagram. The \citet{2017ApJ...836..182J} AGN-selection criterion was used to select, among the sources with the lowest proper motions, $\num{2600}$ QC, $\num{1608}$ of which have a previous label in our catalogue. In this way, and as shown in Fig.~\ref{fig:QC_EMU_PS_conf_matrix}, the quasar selection by \citet{2021PASA...38...46N} has a precision of $\mathcal{P}^{\mathrm{AGN}} = \qty{99.8}{\percent}$ and a recall of $\mathcal{R}^{\mathrm{AGN}} = \qty{76.3}{\percent}$. The differences in metrics are expected, as their method is optimised for precision rather than for a balance between precision and recall.

    \subsection{Predicted radio luminosities}\label{sec:predicted_lums}

\begin{figure}[t]
        \centering
        \begin{overpic}[width=0.99\linewidth, percent]{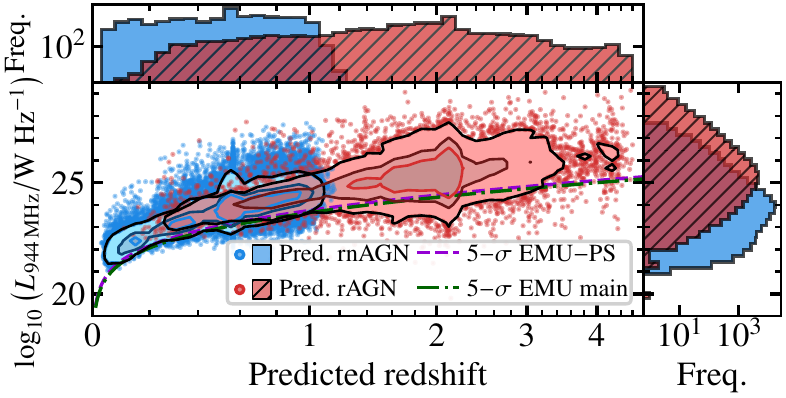}
            \put(12, 47){(a)\phantomsubcaption\label{fig:histogram_log_lum_EMU_PS}}
            \put(12, 36){(b)\phantomsubcaption\label{fig:log_lum_vs_pred_z_EMU_PS}}
            \put(94, 36){(c)\phantomsubcaption\label{fig:histogram_pred_z_EMU_PS}}
        \end{overpic}
        \caption{Distribution of (a) predicted redshifts and (c) $\qty{944}{\MHz}$ luminosities for predicted non-AGNs (rnAGN; in blue) and predicted AGNs in our sample (in hatched red)\@. (b) Two-dimensional distribution of predicted photometric redshifts versus EMU-PS luminosities of predicted non-AGNs in blue and predicted AGNs in red. Contours contain $\qtylist[list-separator = {,\ \linebreak[0]}]{50; 75; 95}{\percent}$ of the sources and coloured points are outliers. Dashed purple and dotted-dashed green lines show the $\qty{5}{\sigma}$ detection limit from the EMU-PS \citep{2021PASA...38...46N} and the main EMU \citep{2025PASA...42...71H} surveys, respectively, assuming a fixed spectral index of $\alpha = -0.7$.}\label{fig:distribution_log_lum_vs_pred_z_EMU_PS}
    \end{figure}

    We derived radio luminosities of selected AGNs and non-AGNs using predicted photometric redshifts and in-band spectral indices. In Fig.~\ref{fig:distribution_log_lum_vs_pred_z_EMU_PS}, we present the distribution of EMU luminosities versus predicted redshifts, with number counts peaking at $z \approx \num{0.58}$ for radio non-AGNs and $z \approx \num{1.2}$ for radio AGNs\@.
    
    Predicted radio non-AGNs peak at $L_{\qty{944}{MHz}} \sim \qty{e24}{\W.\Hz^{-1}}$, with a large high-luminosity tail, challenging the accepted view that high radio powers are necessarily AGN driven \citep[e.g.][]{2014MNRAS.442..682M, 2024A&A...685A..79W}. Although a more detailed investigation is beyond the scope of this work, we suggest two possible explanations. First, some of these sources may have large redshift prediction errors (Appendix~\ref{app:catastrophic_non_agn}). Second, many radio-bright non-AGNs in our sample may have multi-wavelength photometry compatible with, for example, HERGs, which show SF signatures in optical bands while hosting a powerful radio AGNs \citep[e.g.][]{2023A&A...679A..76D, 2024PASA...41...16P, 2025MNRAS.539..463C}. This behaviour highlights the importance of using as much multi-wavelength information as possible and motivates more nuanced indicators such as $p(\mathrm{AGN})$ or those from colour criteria and SED fitting techniques, which can incorporate a wider range of source classes \citep{2026PASA...43...47P}.

    Additionally, we highlight the sudden break in the redshift distribution for non-AGNs at $z \sim \num{1.2}$, also seen in Figs.~\ref{fig:large_pred_z_vs_class_prob} and~\ref{fig:EMU_PS_true_pred_z}. This drop goes against the known presence of radio-bright non-AGNs, mostly of SF galaxies, at high redshifts \citep[$z \gtrsim \num{2}$, e.g.][]{2020ApJ...903..138A, 2024MNRAS.532.1504J, 2025MNRAS.539.1856H}. Since the bulk of the training data for the redshift prediction only reaches $z \sim 1.2$ (see Appendix~\ref{app:distribution_broadening}), the model is forced to place predicted non-AGNs within such a range. One way to solve this training bias is through the inclusion of more high-redshift non-AGNs in the training sample from deeper or wider surveys.

    Furthermore, AGN luminosities show a peak at $\log_{10} (L_{\qty{944}{MHz}} / \unit{\W.\Hz^{-1}})\sim\num{25}$, which is consistent with them being typically brighter than non-AGNs in the radio. At the lower-$z$ end, AGN distribution finishes smoothly as it is compatible with the original EMU-PS catalogue not including extended sources, which are more likely to be seen at lower redshifts. We took the source with the lowest predicted redshift to test our predictions, \texttt{J202943.67-552247.6}. It is predicted to be at $z = \num{0.107}$ and it has a counterpart at $\qty{16}{\arcsec}$ in the R{\"o}ntgen Satellite (ROSAT) All-Sky Bright Source Catalogue \citep{1999A&A...349..389V}. Additionally, it is located outside the AGN regions of Fig.~\ref{fig:WISE_C23_CM_EMU_PS_pred_sources}. According to \citet{2010AJ....140.1868W} and \citet{2017ApJ...836..182J}, this predicted AGN has the colours of elliptical galaxies, stars, and (although less likely) spiral, passive, and LINER galaxies \citep{2023A&A...679A..76D}.

    This source also has an EMU flux of $\sim\qty{3}{mJy}$ with a spectral index of $\alpha = \num{-0.98}$. Additionally, it has $p(\mathrm{AGN}) \sim \num{0.8}$, showing confidence in its classification. Assuming this source is effectively a galaxy, it matches previous findings on the increasing fraction of early-type galaxies hosting low-luminosity AGNs and of red-sequence galaxies at $z \lesssim \num{0.1}$ \citep{2013MNRAS.429.1827P, 2018MNRAS.479..807L}.

    Another relevant feature in Fig.~\ref{fig:log_lum_vs_pred_z_EMU_PS} is the high density of predicted non-AGNs at $z \approx \num{0.58}$ and predicted AGNs at $z \approx \num{2.2}$, showing a larger luminosity spread. Previous works have found that the peaks of both the co-moving AGN luminosity and SF activity densities occur at $z \sim \num{2}$ \citep{2014ARA&A..52..589H, 2015ApJ...813L...8M, 2022MNRAS.509.3015H, 2022ApJ...924..133R, 2023ApJS..265....5H}. We can only match these values to our AGN, as the non-AGN redshifts are limited by the training sample. We explored a possible explanation for this broadening in Appendix~\ref{app:distribution_broadening}, where we show that magnitude imputation biases predictions towards these specific redshifts for sources with sparse photometry.

    \subsection{EMU in-band spectral index evolution}\label{sec:spectral_indices}

\begin{figure}[t]
        \centering
        \begin{overpic}[width=0.99\linewidth, percent]{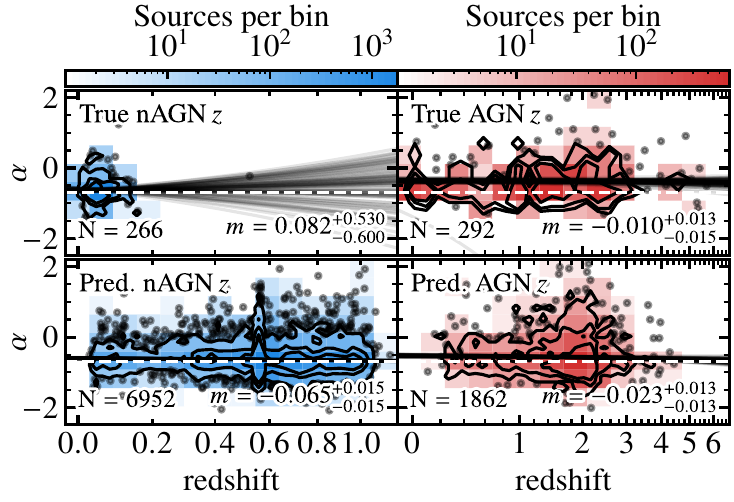}
            \put(49, 52){(a)\phantomsubcaption\label{fig:true_z_vs_spec_index_SFG}}
            \put(93, 52){(b)\phantomsubcaption\label{fig:true_z_vs_spec_index_AGN}}
            \put(49, 29){(c)\phantomsubcaption\label{fig:pred_z_vs_spec_index_SFG}}
            \put(93, 29){(d)\phantomsubcaption\label{fig:pred_z_vs_spec_index_AGN}}
        \end{overpic}
        \caption{Two-dimensional distribution of in-band spectral indices ($\alpha$) versus (predicted or true) redshift of (a) true non-AGNs, (b) true AGNs, (c) predicted non-AGNs, and (d) predicted AGNs with integrated fluxes above $\qty{2.5}{\milli\jansky}$. Contours contain $\qtylist[list-separator = {,\ \linebreak[0]}]{50; 75; 95}{\percent}$ of sources. Black points are outliers. The background plots show the number of sources following colour bars. The number of sources included in each panel is also shown. Black lines show random realisations from the MCMC line fitting posterior between redshift and $\alpha$ (best-fit slope, $m$, in the bottom right corner of each panel, including the $\num{16}$th, $\num{50}$th, and $\num{84}$th percentiles)}\label{fig:pred_z_vs_spec_index}
    \end{figure}

    Radio spectral indices probe the physical processes driving radio emission and can reveal any possible change across cosmic time. Although AGN- and SF-dominated processes often cluster around $\alpha = \num{-0.7}$, which is typical of optically thin synchrotron emission \citep{2013ApJ...768...37C}, this quantity alone does not clearly separate the two populations, motivating an analysis of its evolution with redshift. Thus, we studied the in-band radio spectral indices ($\alpha$) measured by EMU-PS for both predicted populations. Following \citet{2021PASA...38...46N}, we selected sources with fluxes above $\qty{2.5}{\milli\jansky}$ to avoid high $\alpha$ uncertainties. Their evolution, for true and predicted radio AGNs and non-AGNs, is shown in Fig.~\ref{fig:pred_z_vs_spec_index}.

    Spectral indices for true radio-detected non-AGNs (Fig.~\ref{fig:true_z_vs_spec_index_SFG}) have a concentrated distribution that hampers extensive analysis. Nevertheless, all four distributions are roughly symmetric around $\alpha = \num{-0.7}$, which is characteristic of optically thin synchrotron emission from both AGN or non-AGN activity. The latter also coexist with free--free emission from gas ionised by massive stars \citep[$\alpha = \num{-0.1}$;][]{1992ARA&A..30..575C, 2025A&A...703A.202V}. For predicted sources, their dispersion is larger at $z \approx \num{2.2}$ for AGNs and at $z \approx \num{0.58}$ for non-AGNs\@. At these redshifts, some sources show spectral indices as extreme as $\alpha \approx \num{-2.0}$.
    
    To expose any possible correlation between both quantities, we fitted different lines to each subset (predicted full or complete non-AGNs or AGNs) using a Markov chain Monte Carlo Ensemble sampler \citep[MCMC;][]{2010CAMCS...5...65G}. All four distributions show weak, or almost absent, evolution with photometric redshifts, even taking into account the observed frequency shifting from $z = \num{0}$ to $\sim \qty{150}{\MHz}$ at $z = \num{5}$, where additional effects might impact measured $\alpha$ \citep[e.g.][]{2015aska.confE..71A}.

    We compared our results with previous studies. For non-AGNs, we focused on \citet{2017A&A...602A...4D}, which analysed IR- and radio-detected sources up to $z \sim \num{6}$ in the Cosmic Evolution Survey \citep[COSMOS;][]{2007ApJS..172....1S} field. Our distribution of true and predicted spectral indices agrees with theirs, confirming the departure from models that use simple assumptions on the origin and interaction between radio emission sources in SFGs.

    More recently, \citet{2025MNRAS.537.3481P} analysed the redshift evolution of spectral indices for compact non-AGNs and AGNs in radio observations at $\qtylist[list-units = single, list-separator = {,\ \linebreak[0]}]{144;390;610;688;1280}{\MHz}$ over the area of the \textit{XMM-Newton} Large-Scale Structure \citep[\textit{XMM}-LSS;][]{2004JCAP...09..011P} survey. The authors determined that, contrary to previous works \citep[e.g.][]{2000A&AS..143..303D}, using information from both bright and faint radio detections leads to weak or absent connection between $\alpha$ and $z$. Our results provide further support for this picture; that is, when faint radio sources are taken into account, spectral indices no longer show a clear evolution with redshift.

    \subsection{Impact of data quality on predictions}

    Predictions from the pipeline depend on the coverage and quality of both training and inference data \citep{2023PASA...40....1H}. The redshift range of the training data directly constrains the distribution of predicted photometric redshifts. As shown in Appendix~\ref{app:distribution_broadening}, the model does not predict redshifts beyond approximately $z = \num{2}$ for non-AGNs and $z = \num{5}$ for AGNs\@. This behaviour illustrates the importance of an adequate coverage of the parameter space.

    The EMU-PS input data introduce additional sources of uncertainty. In particular, a high fraction of imputed magnitudes can modify the effective representation of a source. Larger imputation fractions can lead to less reliable photometric redshifts, as shown in Appendices~\ref{app:catastrophic_non_agn} and~\ref{app:distribution_broadening}. The cross-matching procedure can have a similar effect if measurements from different sources are incorrectly associated with a single CatWISE source, thus modifying the input properties supplied to the models. Following the analysis of \citet{2013MNRAS.429.2080W} and \citet{2021PASA...38...46N}, Appendix~\ref{app:search_radius} investigates the optimal cross-matching radii between CatWISE and the other catalogues used in this work.

    \subsection{Extended EMU-PS catalogue}

    We compiled the results from the application of the prediction pipeline over the CatWISE-detected sources of the EMU-PS area into an extension of their original catalogue. Apart from source identification variables (including initial classifications and redshift, when available), we appended the predicted classification (together with their probabilities) and photometric redshift. Additionally, we included radio measurements from EMU (fluxes, spectral indices, and deconvolved sizes) as well as optical-to-infrared magnitudes. In Appendix~\ref{app:catalogue_columns}, a detailed description of the columns of the extended catalogue is presented.

    \subsection{Expectations for main EMU survey source selection}

    Finally, to assess the potential of the \citetalias{2023A&A...679A.101C} pipeline for the ongoing full EMU survey, we obtained its expected AGN fraction. The EMU-PS catalogue of compact sources contains $\num{178921}$ radio detections, with $\num{129361}$ of them having non-stellar CatWISE counterparts. Among these, $\num{30866}$ CatWISE-EMU detections ($\qty{17}{\percent}$ of all detections) are predicted to be AGNs\@. EMU is expected to detect $\sim \num{20}$ million extragalactic sources \citep[estimate from the Tiered Radio Extragalactic Continuum Simulation, T-RECS;][]{2019MNRAS.482....2B, 2025PASA...42...71H}. Applied to the main EMU survey, covering $\qty{2 \pi}{\steradian}$, the \citetalias{2023A&A...679A.101C} pipeline could select more than $\num{3.4}$ million compact AGN candidates (or $\sim \num{170}$ compact AGNs per square degree) at $z \lessapprox \num{5}$, with precision and recall above $\qty{95}{\percent}$.

\section{Summary and conclusions}\label{sec:section_06}

    We applied the \citetalias{2023A&A...679A.101C} and Carvajal et al.\ (in prep.) multi-stage ML pipeline to MIR- and radio-detected sources in the EMU-PS area. This pipeline identifies the main origin of emission and radio detectability of extragalactic sources, as well as estimates their photometric redshifts, using optical-to-IR photometry. As source radio detectability is given by the EMU-PS catalogue, we bypassed the second pipeline stage. The resulting classifications and photometric redshifts were compiled into an extension of the EMU-PS value-added source catalogue of compact sources presented by \citet{2021PASA...38...46N}, as described in Appendix~\ref{app:catalogue_columns}.

    The pipeline selected $\num{30866}$ AGN and $\num{98495}$ non-AGN candidates; this is an increase of, respectively, $\sim \num{15}$ and $\sim \num{100}$ times the number of previously confirmed sources in the EMU-PS area. Prediction metrics suggest that $\qty{96.6}{\percent}$ of predicted AGNs and $\qty{92.4}{\percent}$ of predicted non-AGNs are true extragalactic sources. We also recovered $\qty{96.4}{\percent}$ of known radio AGNs and $\qty{93.7}{\percent}$ of known radio non-AGNs in the area. The pipeline also identifies more AGN candidates than the original EMU-PS quasar selection.

    When placed on known AGN selection colour diagrams, previously known EMU-PS sources fit adequately within the regions defined by \citetalias{2012MNRAS.426.3271M}, \citetalias{2018MNRAS.478.3056B}, and \citetalias{2023A&A...679A.101C}, whereas many new candidates extend beyond their expected loci \citep[see also][]{2026MNRAS.tmp..412W}. This behaviour is expected as the pipeline uses a broader set of properties than colour cuts alone \citep{2024ApJ...971...68W}.

    The predicted populations also show distinct radio behaviours. A non-negligible fraction of predicted low-redshift non-AGNs exhibit luminosities and properties more typical of AGNs, revealing overlap between photometric and radio classification. Meanwhile, in-band EMU-PS spectral indices show no clear evolution with predicted redshift for AGNs, which is consistent with previous studies of both faint and bright sources. Predicted radio luminosities also follow established trends.

    Overall, the ML-based pipeline produces a large catalogue of EMU-PS sources spanning a wide range of radio properties. While individual predictions may have uncertainties, the full sample enables population studies of AGN- and SF-dominated sources that extend beyond traditional colour-based selection. This approach will accelerate the classification and characterisation of extragalactic populations in the full EMU survey and future SKA observations.

\section*{Data availability}

    The resulting classifications and photometric redshifts of EMU-PS sources are available in electronic form at the CDS via anonymous ftp to cdsarc.u-strasbg.fr (130.79.128.5) or via \url{http://cdsweb.u-strasbg.fr/cgi-bin/qcat?J/A+A/}. A visual representation of the sources and some of their properties is also available in the web dashboard at \url{https://racarvajalp-emups-ml-dashboard.hf.space/app}.

\begin{acknowledgements}
    The authors would like to thank Adeel Ahmad, Ian Smail, Mattia Vaccari, Sarah White, and Christian Wolf, as well as the anonymous referee, for their insightful suggestions and comments on the manuscript.
    
    This work was supported by the Portuguese Funda\c{c}\~{a}o para a Ci\^{e}ncia e a Tecnologia (FCT) through research grants PTDC\slash{}FIS-AST\slash{}29245\slash{}2017, EXPL\slash{}FIS-AST\slash{}1085\slash{}2021 (\textsc{doi}: \href{https://doi.org/10.54499/EXPL/FIS-AST/1085/2021}{10.54499\slash{}EXPL\slash{}FIS-AST\slash{}1085/2021}), UID\slash{}FIS\slash{}04434\slash{}2019, UIDB\slash{}04434\slash{}2020 (\textsc{doi}: \href{https://doi.org/10.54499/UIDB/04434/2020}{10.54499\slash{}UIDB\slash{}04434\slash{}2020}), UIDP\slash{}04434\slash{}2020 (\textsc{doi}: \href{https://doi.org/10.54499/UIDP/04434/2020}{10.54499\slash{}UIDP\slash{}04434\slash{}2020}), and UID\slash{}04434\slash{}2025 (\textsc{doi}: \href{https://doi.org/10.54499/UID/04434/2025}{10.54499\slash{}UID\slash{}04434\slash{}2025}). R.C. also acknowledges support from the Funda\c{c}\~{a}o para a Ci\^{e}ncia e a Tecnologia (FCT) through the Fellowship PD\slash{}BD\slash{}150455\slash{}2019 (PhD::SPACE Doctoral Network PD\slash{}00040\slash{}2012) and POCH\slash{}FSE (EC) as well as from the GEELSBE2 project with reference PID2023-150393NB-I00 funded by MCIU\slash{}AEI\slash{}10.13039\slash{}501100011033 and the FSE+, and also the Consolidaci\'{o}n Investigadora IGADLE project with reference CNS2024-154572. R.C. also gratefully acknowledges financial support of the Department of Education, Junta de Castilla y Le\'{o}n and FEDER Funds (Reference: CLU-2023-1-05). C.M.P. acknowledges funding from a UKRI Future Leaders Fellowship (grant codes MR\slash{}T020989\slash{}1 and MR\slash{}Y019539\slash{}1).

    This scientific work uses data obtained from Inyarrimanha Ilgari Bundara / the CSIRO Murchison Radio-astronomy Observatory. We acknowledge the Wajarri Yamaji People as the Traditional Owners and native title holders of the Observatory site. CSIRO's ASKAP radio telescope is part of the Australia Telescope National Facility (\url{https://ror.org/05qajvd42}). Operation of ASKAP is funded by the Australian Government with support from the National Collaborative Research Infrastructure Strategy. ASKAP uses the resources of the Pawsey Supercomputing Research Centre. Establishment of ASKAP, Inyarrimanha Ilgari Bundara, the CSIRO Murchison Radio-astronomy Observatory and the Pawsey Supercomputing Research Centre are initiatives of the Australian Government, with support from the Government of Western Australia and the Science and Industry Endowment Fund.

    This publication makes use of data products from the Wide-field Infrared Survey Explorer, which is a joint project of the University of California, Los Angeles, and the Jet Propulsion Laboratory/California Institute of Technology, funded by the National Aeronautics and Space Administration.
    
    This publication makes use of data products from the Two Micron All Sky Survey, which is a joint project of the University of Massachusetts and the Infrared Processing and Analysis Center/California Institute of Technology, funded by the National Aeronautics and Space Administration and the National Science Foundation.

    This project used public archival data from the Dark Energy Survey (DES). Funding for the DES Projects has been provided by the U.S. Department of Energy, the U.S. National Science Foundation, the Ministry of Science and Education of Spain, the Science and Technology Facilities Council of the United Kingdom, the Higher Education Funding Council for England, the National Center for Supercomputing Applications at the University of Illinois at Urbana-Champaign, the Kavli Institute of Cosmological Physics at the University of Chicago, the Center for Cosmology and Astro-Particle Physics at the Ohio State University, the Mitchell Institute for Fundamental Physics and Astronomy at Texas A\&M University, Financiadora de Estudos e Projetos, Funda{\c c}{\~a}o Carlos Chagas Filho de Amparo {\`a} Pesquisa do Estado do Rio de Janeiro, Conselho Nacional de Desenvolvimento Cient{\'\i}fico e Tecnol{\'o}gico and the Minist{\'e}rio da Ci{\^e}ncia, Tecnologia e Inova{\c c}{\~a}o, the Deutsche Forschungsgemeinschaft, and the Collaborating Institutions in the Dark Energy Survey. The Collaborating Institutions are Argonne National Laboratory, the University of California at Santa Cruz, the University of Cambridge, Centro de Investigaciones Energ{\'e}ticas, Medioambientales y Tecnol{\'o}gicas-Madrid, the University of Chicago, University College London, the DES-Brazil Consortium, the University of Edinburgh, the Eidgen{\"o}ssische Technische Hochschule (ETH) Z{\"u}rich,  Fermi National Accelerator Laboratory, the University of Illinois at Urbana-Champaign, the Institut de Ci{\`e}ncies de l'Espai (IEEC/CSIC), the Institut de F{\'\i}sica d'Altes Energies, Lawrence Berkeley National Laboratory, the Ludwig-Maximilians Universit{\"a}t M{\"u}nchen and the associated Excellence Cluster Universe, the University of Michigan, the National Optical Astronomy Observatory, the University of Nottingham, The Ohio State University, the OzDES Membership Consortium, the University of Pennsylvania, the University of Portsmouth, SLAC National Accelerator Laboratory, Stanford University, the University of Sussex, and Texas A\&M University. 
    Based in part on observations at Cerro Tololo Inter-American Observatory, National Optical Astronomy Observatory, which is operated by the Association of Universities for Research in Astronomy (AURA) under a cooperative agreement with the National Science Foundation.

    This work has made use of data from the European Space Agency (ESA) mission \textit{Gaia} (\url{https://www.cosmos.esa.int/gaia}), processed by the \textit{Gaia} Data Processing and Analysis Consortium (DPAC, \url{https://www.cosmos.esa.int/web/gaia/dpac/consortium}). Funding for the DPAC has been provided by national institutions, in particular the institutions participating in the \textit{Gaia} Multilateral Agreement.

    This research has made use of the Astrophysics Data System, funded by NASA under Cooperative Agreement 80NSSC21M0056, TOPCAT \citep{2005ASPC..347...29T}, and \texttt{JupyterLab} (\texttt{v4.2.4}).
    
    It also made extensive use of the VizieR catalogue access tool \citep[][\textsc{doi}: \href{https://doi.org/10.26093/cds/vizier}{10.26093/cds/vizier}]{vizier} as well as the CDS cross-match service \citep{2012ASPC..461..291B, 2020ASPC..522..125P}, the SIMBAD database \citep{2000A&AS..143....9W}, and \texttt{hips2fits},\footnote{\url{https://alasky.cds.unistra.fr/hips-image-services/hips2fits}}, all services provided by CDS, Strasbourg Astronomical Observatory, France.

    This work made extensive use of the \texttt{Python} packages \texttt{Astropy}\footnote{\url{https://www.astropy.org}} \citep[\texttt{v5.0};][]{astropy:2013, astropy:2018, 2022ApJ...935..167A}, \texttt{Astroquery}\footnote{\url{https://github.com/astropy/astroquery}} \citep[\texttt{v0.4.8};][]{2019AJ....157...98G}, \texttt{CMasher}\footnote{\url{https://github.com/1313e/CMasher}} \citep[\texttt{v1.6.3};][]{2020JOSS....5.2004V}, \texttt{Colorcet}\footnote{\url{https://colorcet.holoviz.org}} \citep[\texttt{v3.1.0};][]{glasbey_2007, 2015arXiv150903700K}, \texttt{emcee}\footnote{\url{https://github.com/dfm/emcee}} \citep[\texttt{v3.1.6};][]{2013PASP..125..306F}, \texttt{hips}\footnote{\url{https://github.com/hipspy/hips}} \citep[\texttt{v0.2};][]{2015A&A...578A.114F}, \texttt{Matplotlib} \citep[\texttt{v3.5.1};][]{Hunter:2007}, \texttt{MetBrewer}\footnote{\url{https://github.com/BlakeRMills/MetBrewer}}, \texttt{NumPy} \citep[\texttt{v1.20.3};][]{harris2020array}, \texttt{pandas} \citep[\texttt{v1.4.2};][]{pandas}, \texttt{panel}\footnote{\url{https://panel.holoviz.org}} (\texttt{v1.4.5}), \texttt{PyCCL}\footnote{\url{https://github.com/LSSTDESC/CCL}} \citep[\texttt{v2.6.1};][]{2019ApJS..242....2C}, \texttt{scikit-learn} \citep[\texttt{v0.23.2};][]{scikit-learn}, and \texttt{SciPy} \citep[\texttt{v1.5.3};][]{2020SciPy-NMeth}.
\end{acknowledgements}

\bibliographystyle{aa}
\bibliography{aa60860-26corr}

\begin{appendix}

\section{Redshift prediction confidence intervals}\label{app:redshift_intervals}

    As mentioned in Sect.~\ref{sec:section_02}, the redshift models in the prediction pipeline do not output uncertainty estimates. We therefore used conformal prediction \citep{vovk2005algorithmic} to construct prediction intervals with a given target marginal coverage of $1 - \alpha_{\text{conf}} = \qty{68}{\percent}$ ($\qty{1}{\sigma}$). Following the normalised split conformal framework \citep{Papadopoulos2002} and the redshift-specific procedure by \citet{2026AJ....171..114S}, we first trained an auxiliary random forest model \citep{2001MachL..45....5B} on the non-catastrophic labelled sources ($\vert \Delta z \vert / (1 + z_{\text{true}}) \leq \num{0.15}$) using an out-of-fold cross-validation scheme to predict the expected absolute redshift scatter, $\vert \Delta z \vert$, as a function of $z_{\text{pred}}$ and $p(\text{AGN})$.
    
    Residuals, $\Delta z$, are then normalised by the predicted scatter and separate calibration factors, $q_{\text{lo}}$ and $q_{\text{hi}}$, are obtained as the $\lceil (n+1)(1-\alpha_{\text{conf}}/2) \rceil / n$ quantiles of the absolute values of the negative and positive normalised residuals \citep[a finite-sample corrected version of $\num{1} - \alpha_{\text{conf}}/2$;][]{AngelopoulosIntro2023}. This step produces asymmetric intervals $[z_{\text{pred}} - q_{\text{lo}}\hat{\sigma},\, z_{\text{pred}} + q_{\text{hi}}\hat{\sigma}]$, with $\hat{\sigma}$ the predicted absolute scatter and the lower bounds clipped at $z = \num{0}$.
    
    Finally, after the conformal quantiles were calibrated, a separate random forest model for the scatter, trained on the $z_{\text{pred}}$ and $p(\text{AGN})$ values of all non-catastrophic labelled sources, was applied to the full EMU-PS catalogue. This model produces the source-specific redshift prediction intervals with empirically verified coverage of $\qty{84}{\percent}$ (more conservative than the $\qty{68}{\percent}$ target) that are included in the final catalogue (Appendix~\ref{app:catalogue_columns}).

\section{Counterpart search radius determination}\label{app:search_radius}

\begin{figure}[ht!]
        \centering
        \begin{overpic}[width=0.99\linewidth, percent]{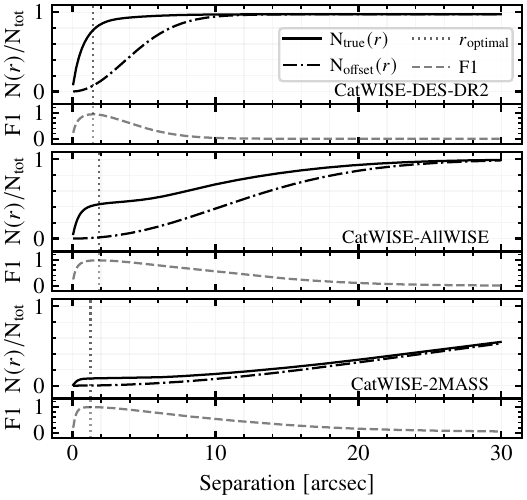}
            \put(12, 89){(a)\phantomsubcaption\label{fig:match_fraction_CatWISE_DES}}
            \put(12, 61){(b)\phantomsubcaption\label{fig:match_fraction_CatWISE_AllWISE}}
            \put(12, 33){(c)\phantomsubcaption\label{fig:match_fraction_CatWISE_2MASS}}
        \end{overpic}
        \caption{Cumulative cross-match distribution for sources in CatWISE and (a) DES-DR2, (b) AllWISE, and (c) 2MASS\@ normalised by the total number of CatWISE sources in the EMU-PS area, $\text{N}_\text{tot}$. Solid curves show original matches, $\text{N}_\text{true}(r)$, dot-dashed curves the $\qty{1}{\arcminute}$-offset matches, $\text{N}_\text{offset}(r)$, and dotted vertical lines show optimal search radii. Lower panels show the F1 score as a function of separation.}\label{fig:match_fraction_CatWISE_catalogues}
    \end{figure}

    To optimise the counterpart selection (Sect.~\ref{sec:section_02}), we cross-matched CatWISE, independently, against DES-DR2, AllWISE, and 2MASS, using TOPCAT \citep{2005ASPC..347...29T} symmetric sky matching out to $\qty{30}{\arcsec}$. To estimate contamination by chance associations, we repeated each match after shifting CatWISE coordinates by $\qty{1}{\arcminute}$ in an arbitrary direction \citep[e.g.][]{2013MNRAS.429.2080W, 2021PASA...38...46N}. This offset removes genuine physical associations while preserving source density structure. The cumulative numbers of real, $\text{N}_\text{true}(r)$, and offset, $\text{N}_\text{offset}(r)$, matches are shown in Fig.~\ref{fig:match_fraction_CatWISE_catalogues}.

    These curves map directly to the elements of a confusion matrix. $\text{N}_\text{true}(r)$ is the combined count of true and false positives (i.e.\ genuine and chance matches within $r$), while $\text{N}_\text{offset}(r)$ estimates the false positives alone. The excess $\text{N}_\text{true}(r) - \text{N}_\text{offset}(r)$ is therefore, an empirical estimator of the true positives recovered within $r$, and its maximum over $r$ estimates the total number of genuine associations \citep{2021PASA...38...41F}. Therefore, precision \citep[also called reliability in this context;][]{1975AJ.....80..887C} and recall (Eqs.~\ref{eq:precision} and~\ref{eq:recall}) are obtained through
    \begin{eqnarray}\label{eq:prec_recall_match}
        \mathcal{P}(r) &=& \dfrac{\text{N}_\text{true}(r) - \text{N}_\text{offset}(r)}{\text{N}_\text{true}(r)} = 1 - \dfrac{\text{N}_\text{offset}(r)}{\text{N}_\text{true}(r)}\,,\\
        \mathcal{R}(r) &=& \dfrac{\text{N}_\text{true}(r) - \text{N}_\text{offset}(r)}{\max\limits_{r}{\{\text{N}_\text{true}(r) - \text{N}_\text{offset}(r)\}}}\,,
    \end{eqnarray}
    \noindent from which the F1 score follows using Eq.~(\ref{eq:f1_score}). We adopt, for each catalogue pair, the radius that maximises F1 (vertical dotted lines in Fig.~\ref{fig:match_fraction_CatWISE_catalogues}): $r \approx \ang[angle-symbol-over-decimal]{;;1.5}$ for DES-DR2, $r \approx \ang[angle-symbol-over-decimal]{;;1.9}$ for AllWISE, and $r \approx \ang[angle-symbol-over-decimal]{;;1.3}$ for 2MASS\@.

\section{Imputation in selected bands}\label{app:imputation_metrics}

    Table~\ref{table:non_imputed_sources} summarises the degree of imputation for each band. We list the absolute number of non-star, non-imputed sources, and the percentage over the number of CatWISE-detected sources ($\num{8291846}$) and those with a detection in the EMU-PS catalogue of compact sources (CW+EMU-PS, $\num{129361}$). Even for CatWISE bands, the number of imputed sources is non-zero because the catalogue requires a $\qty{5}{\sigma}$ detection in only one of two bands for inclusion \citep{2021ApJS..253....8M}.

    \begin{table}[ht!]
        \setlength{\tabcolsep}{3pt}
        \centering
        \caption[Measured magnitudes]{Number and percentage of non-imputed sources for bands in the EMU-PS area over base and radio-detected samples.}\label{table:non_imputed_sources}%
        \begin{tabular}{c c S[table-format=8] S[table-format=3.1] S[table-format=6] S[table-format=3.1]}
        \toprule
        {Survey}                 & {Band}      & \multicolumn{4}{c}{Non-imputed sources number}  \\
        \cmidrule(lr){3-6}
                                 &             & \multicolumn{2}{c}{CW} & \multicolumn{2}{c}{CW+EMU-PS} \\
        \cmidrule(lr){3-4}\cmidrule(lr){5-6}
                                 &             &          & {(\%)} &        & {(\%)} \\
        \midrule
        \multirow{5}{*}{DES-DR2} & \textit{g}  &  5758730 &  69.5  & 103166 &  79.8 \\
                                 & \textit{r}  &  6016891 &  72,6  & 105977 &  81.9 \\
                                 & \textit{i}  &  6048883 &  73.0  & 106317 &  82.2 \\
                                 & \textit{z}  &  6050559 &  73.0  & 106343 &  82.2 \\
                                 & \textit{y}  &  5813586 &  70.1  & 103284 &  79.8 \\
        \cmidrule(lr){2-6}
        \multirow{3}{*}{2MASS}   & \textit{J}  &    48940 &   0.6  &   9671 &   7.5 \\
                                 & \textit{H}  &    48940 &   0.6  &   9671 &   7.5 \\
                                 & \textit{Ks} &    48940 &   0.6  &   9671 &   7.5 \\
        \cmidrule(lr){2-6}
        \multirow{2}{*}{CatWISE} & \textit{W1} &  8291840 & 100.0  & 129361 & 100.0 \\
                                 & \textit{W2} &  8291846 & 100.0  & 129361 & 100.0 \\
        \cmidrule(lr){2-6}
        \multirow{2}{*}{AllWISE} & \textit{W3} &  2869458 &  34.6  &  90590 &  70.0 \\
                                 & \textit{W4} &  2869991 &  34.6  &  90623 &  70.1 \\
        \bottomrule
        \end{tabular}
        \end{table}

\section{Additional prediction metrics}\label{app:more_prediction_metrics}

    Beyond the metrics presented in Sect.~\ref{sec:pred_metrics}, additional quantities can be used to assess the quality of ML-based predictions. For classification, precision and recall can be combined into the $\text{F}1$ score \citep{Dice_1945, van1979information}, defined as the harmonic mean of both quantities or, using the elements of the confusion matrix as
    \begin{equation}\label{eq:f1_score}
        \text{F}1 = \dfrac{2}{\mathcal{R}^{-1} + \mathcal{P}^{-1}} = \dfrac{2\text{TP}}{2\text{TP} + \text{FN} + \text{FP}} \,.
    \end{equation}
    $\text{F}1$ score ranges from $0$, when all predictions are incorrect, to $1$, when all predictions are correct. For the first, AGN-related step of the pipeline over our sample, we find $\text{F}1^{\text{AGN}} = \qty{96.5}{\percent}$, $\text{F}1^{\text{non-AGN}} = \qty{92.6}{\percent}$, and $\text{F}1^{c} = \qty{93.8}{\percent}$.

    As done in the main text, we compared these values with those by \citet{2023A&A...675A.159K}, whose authors obtain $\text{F}1^{\text{AGN}} = \qty{94}{\percent}$ and $\text{F}1^{\text{non-AGN}} = \qty{83}{\percent}$. They are below our results, reflecting the positive effect of threshold optimisation in the classification (Sect.~\ref{sec:section_02}).

    For regression, we also considered the bias \citep[$b$;][]{2000ApJ...538...29C}, defined as the mean normalised difference between predicted and true redshifts,
    \begin{equation}\label{eq:bias_z}
    b = \left\langle \dfrac{\Delta z}{1 + z_{\text{true}}} \right\rangle \,.
    \end{equation}
    \noindent Smaller absolute bias values indicate better overall performance and its sign reveals whether predictions tend to under- or overestimate redshifts. In our case, $b^{\text{AGN}} = \num{0.009}$, $b^{\text{non-AGN}} = \num{0.022}$, and $b^{c} = \num{0.013}$, showing a mild systematic underprediction of redshifts. These results can be directly compared with those by \citet{2025ApJS..277...41K}, whose authors obtain $b^{c}=\num{-0.01}$, a similar absolute value to ours.

\section{Modified classification threshold}\label{app:new_agn_threshold}

\begin{figure}[ht!]
        \centering
        \begin{overpic}[width=0.99\linewidth, percent]{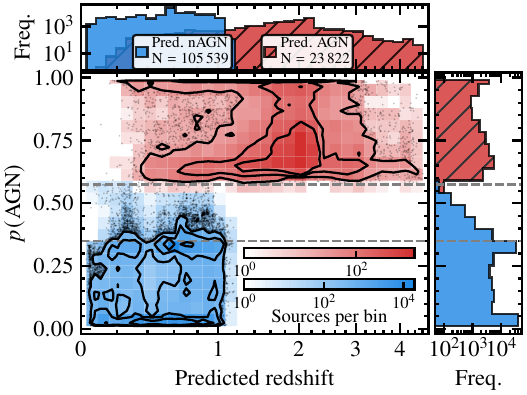}
            \put(76, 70){(a)\phantomsubcaption\label{fig:EMU_PS_all_pred_z_new_thresh}}
            \put(76, 56){(b)\phantomsubcaption\label{fig:pred_z_vs_class_prob_new_thresh}}
            \put(94, 56){(c)\phantomsubcaption\label{fig:EMU_PS_all_class_prob_new_thresh}}
        \end{overpic}
        \caption{Distribution of predicted (a) redshifts and (c) class probabilities for predicted non-AGNs (nAGNs, blue) and predicted AGNs (red hatched). Description as in Fig.~\ref{fig:large_pred_z_vs_class_prob}. Grey dashed lines show probability threshold from \citetalias{2023A&A...679A.101C} and ad-hoc value.}\label{fig:large_pred_z_vs_class_prob_new_thresh}
    \end{figure}

    Besides the AGN probability threshold provided by \citetalias{2023A&A...679A.101C}, alternative values can be used. Based on the probabilities distribution in Fig.~\ref{fig:large_pred_z_vs_class_prob}, with a separation between regions of AGN-predicted probabilities, we tested a different threshold value at $p(\mathrm{AGN}) = 0.575$. Such change implies that some sources are subject to different models (i.e.\ redshift prediction for AGNs or non-AGNs) based on their new classification. The new distribution of probabilities and predicted redshifts is shown in Fig.~\ref{fig:large_pred_z_vs_class_prob_new_thresh}.

    Active galactic nucleus probabilities are distributed more uniformly than before. Additionally, non-AGN probabilities do not have a sudden drop near their new threshold and show instead, a smooth decline towards it. We tested the pertinence of using this modified threshold instead of that given by the prediction pipeline. Firstly, we determined prediction metrics derived from it. For the AGN classification, we obtained recalls of $\mathcal{R}^{\mathrm{AGN}} = \qty{96.0}{\percent}$, $\mathcal{R}^{\mathrm{non-AGN}} = \qty{93.2}{\percent}$, and $\mathcal{R}^{c} = \qty{94.1}{\percent}$. Precisions are $\mathcal{P}^{\mathrm{AGN}} = \qty{96.8}{\percent}$, $\mathcal{P}^{\mathrm{non-AGN}} = \qty{91.6}{\percent}$, and $\mathcal{P}^{c} = \qty{93.2}{\percent}$. $\text{F}1$ scores are $\text{F}1^{\text{AGN}} = \qty{96.4}{\percent}$, $\text{F}1^{\text{non-AGN}} = \qty{92.4}{\percent}$, and $\text{F}1^{c} = \qty{93.7}{\percent}$. Also, cross-entropy loss, that does not depend on class thresholds, was kept at $\mathcal{H} = \num{0.148}$. Then, for the redshift predictions, we retrieved biases of $b^{\text{AGN}} = \num{0.009}$, $b^{\text{non-AGN}} = \num{0.022}$, and $b^{c} = \num{-0.013}$. NMADs are $\sigma_{\text{NMAD}}^{\mathrm{AGN}} = \num{0.103}$, $\sigma_{\text{NMAD}}^{\mathrm{non-AGN}} = \num{0.030}$, and $\sigma_{\text{NMAD}}^{c} = \num{0.065}$. Finally, redshift outlier fractions are $\eta^{\mathrm{AGN}} = \qty{30.8}{\percent}$, $\eta^{\mathrm{non-AGN}} = \qty{1.0}{\percent}$, and $\eta^{c} = \qty{21.5}{\percent}$.

    Some metrics show a slight improvement from those using the pre-defined threshold. However, the focus of this work is not to explore strategies for the selection of optimal probability separations. For this reason, and given it has been already validated, we kept the value provided by \citetalias{2023A&A...679A.101C} throughout this work.

\section{Colour evolution}\label{app:colour_evol}

\begin{figure}[ht!]
        \centering
        \begin{overpic}[width=0.99\linewidth, percent]{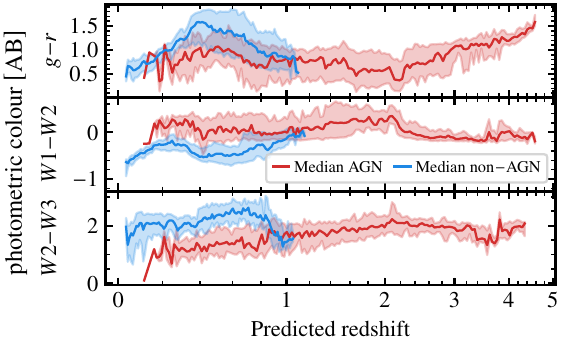}
            \put(22, 57){(a)\phantomsubcaption\label{fig:g_r_redshift_evol}}
            \put(22, 40){(b)\phantomsubcaption\label{fig:W1_W2_redshift_evol}}
            \put(22, 24){(c)\phantomsubcaption\label{fig:W2_W3_redshift_evol}}
        \end{overpic}
        \caption{Median (a) (\textit{g $-$ r}), (b) (\textit{W1 $-$ W2}), and (c) (\textit{W2 $-$ W3}) vs predicted photometric redshift for selected AGNs (red) and non-AGNs (blue). Shaded regions contain values between $\num{16}$th and $\num{84}$th percentiles of the sources within each redshift bin.}\label{fig:colours_redshift_evol}
    \end{figure}

    Expanding the discussion from Sect.~\ref{sec:colour_analysis}, we analysed the evolution of colours with redshift. Figure~\ref{fig:colours_redshift_evol} presents the evolution of (\textit{g $-$ r}), (\textit{W1 $-$ W2}), and (\textit{W2 $-$ W3}) with redshift for our sample.

    Across all colours, AGNs and non-AGNs show distinct distributions, confirming the classification power of the prediction pipeline. For (\textit{g $-$ r}) in Fig.~\ref{fig:g_r_redshift_evol}, AGNs show a clear upturn at $z \gtrsim \num{2}$, coinciding with the onset of Lyman-$\alpha$ forest, used to select high-z sources \citep[e.g.][]{2012ApJS..199....3R, 2017MNRAS.471.3955R}. In Fig.~\ref{fig:W1_W2_redshift_evol}, (\textit{W1 $-$ W2}) follows, for both AGNs and non-AGNs, the behaviour of the templates by \citet{2019MNRAS.489.3351B} and \citet{2022A&A...661A...3S}, with a dip at $z \sim \num{0.8}$. For (\textit{W2 $-$ W3}), in Fig.~\ref{fig:W2_W3_redshift_evol}, non-AGNs are, in general, redder than AGNs, reflecting undiluted SF-related polycyclic aromatic hydrocarbon emission \citep[PAH;][]{2023A&A...679A..76D}.

\section{Non-AGN catastrophic redshifts}\label{app:catastrophic_non_agn}

\begin{figure}[ht!]
        \centering
        \includegraphics[width=0.99\columnwidth]{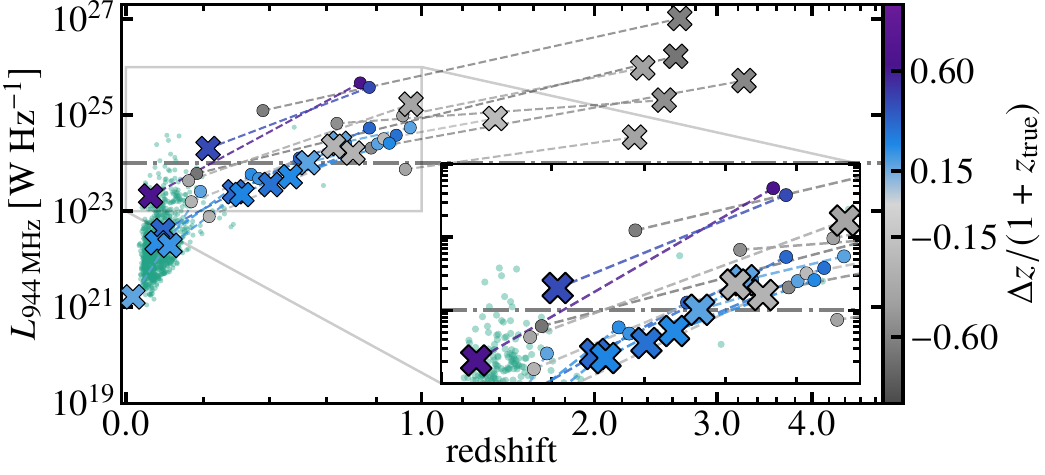}
        \caption{Difference between EMU luminosities calculated using predicted (circles) or true (crosses) redshifts for known sources predicted as non-AGNs with catastrophic redshift errors. Dashed lines connect pairs of points and inset shows detail in the crowded region. Symbols are coloured by redshift dispersion and background green points are predicted non-AGNs with no catastrophic redshift errors. Dashed-dotted line marks $L_{\qty{944}{MHz}} = \qty{e24}{\W.\Hz^{-1}}$.}\label{fig:non_agn_lum_drift}
    \end{figure}

    The outlier fraction for the redshift prediction of correctly classified non-AGNs is $\eta^{\mathrm{non-AGN}} = \qty{1.0}{\percent}$ (Sect.~\ref{sec:selected_sources}). When extended to all predicted non-AGNs with a prior redshift, it increases to $\eta^{\mathrm{non-AGN}} = \qty{2.2}{\percent}$. These values mean that, among the $\num{993}$ known sources predicted as non-AGNs (Fig.~\ref{fig:conf_matrix_SFG}), $\num{22}$ have catastrophic redshift predictions, of which $\num{13}$ are, originally, AGNs\@. Figure~\ref{fig:non_agn_lum_drift} shows the resulting drift in EMU luminosity for these predicted non-AGNs with catastrophic redshifts.

    Additionally, from the $\num{993}$ predicted non-AGNs, $\num{25}$ have estimated luminosities in excess of $L_{\qty{944}{MHz}} = \qty{e24}{\W.\Hz^{-1}}$. Of these, $14$ have catastrophic redshift errors (circles above the dashed-dotted line in Fig.~\ref{fig:non_agn_lum_drift}). Only one true non-AGN has a real luminosity higher than predicted while the others move below the $L_{\qty{944}{MHz}} = \qty{e24}{\W.\Hz^{-1}}$ line (blue-purple points). Thus, $\qty{56}{\percent}$ of apparently too-bright non-AGNs are a product of pipeline artifacts, driven by incorrect class or redshift predictions.

\begin{figure}[ht!]
        \centering
        \includegraphics[width=0.99\columnwidth]{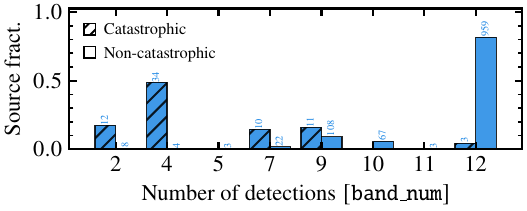}
        \caption{Fractions of predicted non-AGNs with catastrophic (hatched bars) and non-catastrophic (clean bars) redshift predictions, shown as a function of \texttt{band\_num}. Each subsample is normalised by its total size and numbers on top of bars show their absolute counts.}\label{fig:catastrophic_band_num}
    \end{figure}

    We further investigated the connection between magnitude imputation (Sect.~\ref{sec:section_03}) and catastrophic redshift predictions by comparing the number of measured bands, \texttt{band\_num}, for predicted non-AGNs with and without catastrophic redshift errors. The corresponding fractions are shown in Fig.~\ref{fig:catastrophic_band_num}.

    It is possible to see that non-catastrophic predictions are more common in sources detected in nearly all bands. By contrast, catastrophic predictions are concentrated among sources with fewer measured magnitudes. This trend supports a connection between incomplete photometric coverage, its imputation, and the reliability of predicted redshifts.

\section{Predicted redshift broadening}\label{app:distribution_broadening}

    We explored possible explanations for the noticeable broadening of the source distribution seen in Figs.~\ref{fig:distribution_log_lum_vs_pred_z_EMU_PS} and~\ref{fig:pred_z_vs_spec_index}. Predicted non-AGNs show a dispersion in their properties at $z \approx \num{0.58}$ and predicted AGNs at $z \approx \num{2.2}$. We suggest that this effect is mainly due to two factors. The first factor relates to the conditions in which the pipeline models were trained, namely, the training redshifts distribution. The second point is the influence of the magnitude imputation in the predicted sources.

\begin{figure}[ht!]
        \centering
        \includegraphics[width=0.99\columnwidth]{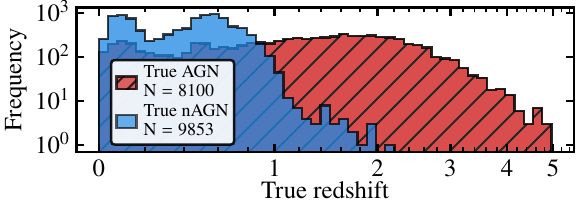}
        \caption{True redshift distribution from pipeline training set. Histograms show true redshifts for, in blue, radio-detected non-AGNs and, in hatched red, true radio-detected AGNs in the LoTSS-DR1 area. Legend shows number of sources in each subset.}\label{fig:HETDEX_DR1_true_z_hist}
    \end{figure}

    For the first aspect, we retrieved the training measurements from the prediction pipeline\footnote{\url{https://zenodo.org/records/10220009}} \citep{carvajal_2023_10220009}. Models were trained with sources in the area of the LoTSS Data Release 1 \citep[LoTSS-DR1;][]{2019A&A...622A...1S}, containing information from a broad range of redshifts. Redshift distributions of radio-detected non-AGNs and AGNs are shown in Fig.~\ref{fig:HETDEX_DR1_true_z_hist}.

    We find two main peaks for non-AGNs\@: one at $z \approx \num{0.1}$ and a broader one at $z \approx \num{0.58}$. Training AGNs have distribution peaks at $z \sim \numlist[list-separator = {,\ \linebreak[0]}]{0.1;0.4;2.2}$. The highest and broadest redshift peaks coincide with the values where large dispersions are seen in predicted sources (see Fig.~\ref{fig:EMU_PS_all_pred_z}), and not with true available values (Fig.~\ref{fig:EMU_PS_true_z}). Thus, we argue that the trained models have more information (i.e.\ cover a broader region of the parameter space) about sources at these training peak redshifts. In other words, trained models learned more about sources around these redshifts and any source that deviates from the learned features has a higher probability of landing close to $z \approx \num{0.58}$ or $z \approx \num{2.2}$ if they are classified as non-AGNs or AGNs, respectively.

\begin{figure}[ht!]
        \centering
        \begin{overpic}[width=0.99\linewidth, percent]{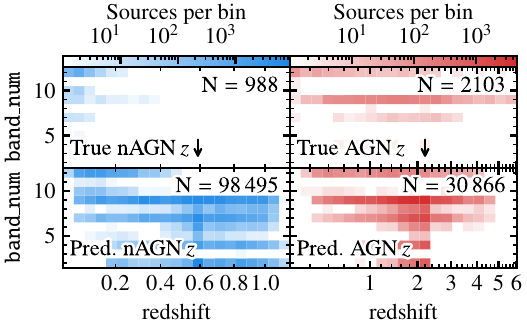}
            \put(49, 33){(a)\phantomsubcaption\label{fig:true_z_vs_band_num_colour_SFG}}
            \put(93, 33){(b)\phantomsubcaption\label{fig:true_z_vs_band_num_colour_AGN}}
            \put(49, 14){(c)\phantomsubcaption\label{fig:pred_z_vs_band_num_colour_SFG}}
            \put(93, 14){(d)\phantomsubcaption\label{fig:pred_z_vs_band_num_colour_AGN}}
        \end{overpic}
        \caption{Two-dimensional distribution of \texttt{band\_num} vs (predicted or true) redshift of (a) true non-AGNs (nAGNs), (b) true AGNs, (c) predicted nAGNs, and (d) predicted AGNs\@. Darker colour implies more sources in a bin, as coded by colour bar. Black arrows point to predicted redshift overdensities.}\label{fig:z_vs_band_num_colour}
    \end{figure}
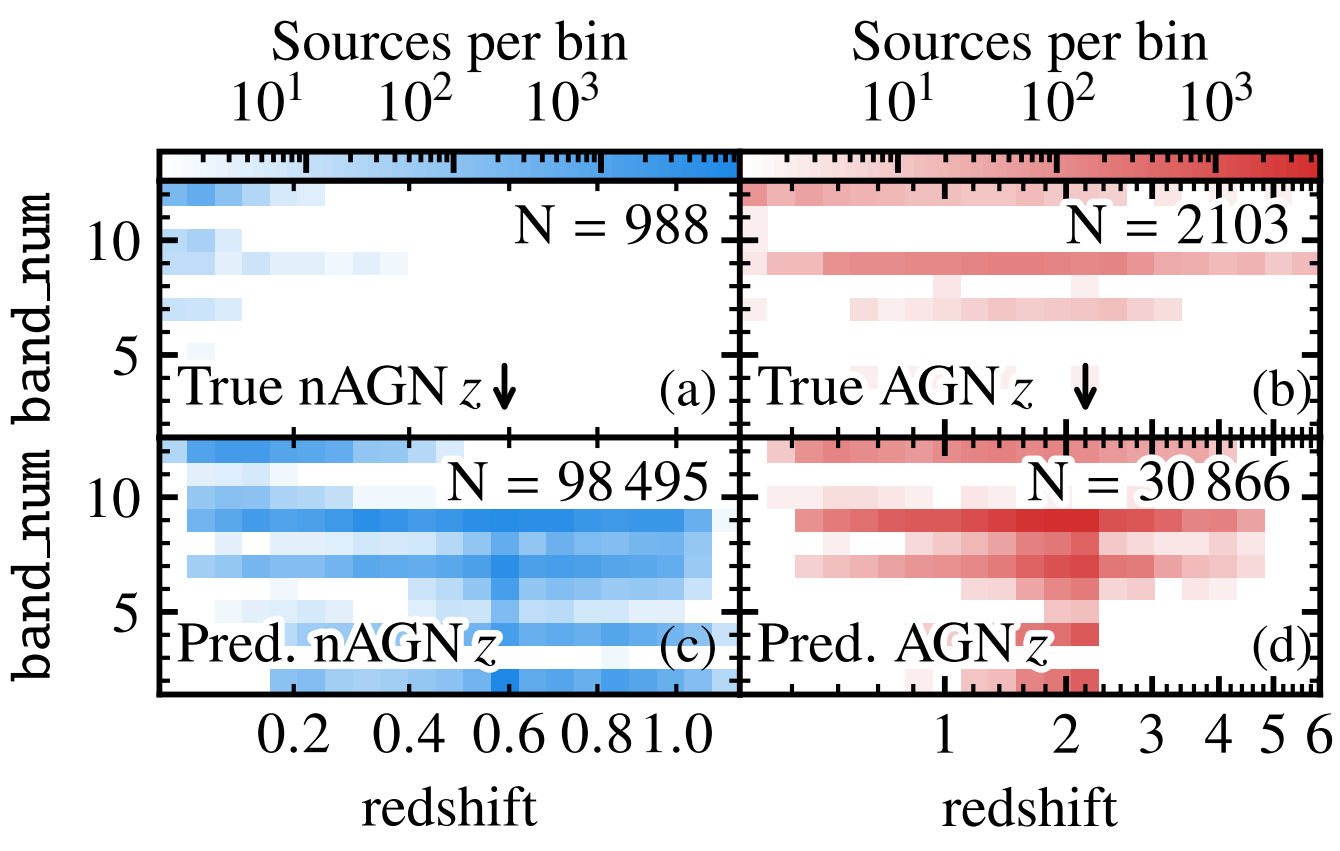

    A second line of evidence comes from magnitude imputations. Sources with missing measurements have the same assigned magnitudes and, consequently, colours. Thus, imputed faint sources have the same input features regardless of their original properties (e.g.\ spectral indices). Such lack of variety in features might overwhelm models and make them prone to predict quantities different from those expected for these sources. One way to see the effect of magnitude imputation is through its dependence with predicted redshift. In Fig.~\ref{fig:z_vs_band_num_colour}, the two-dimensional distributions of the number of measured bands as a function of predicted photometric redshift is shown for known and predicted sources.

    For $\mathtt{band\_num} < \num{10}$, predicted photometric redshifts show two relevant overdense columns. Many non-AGNs have predicted redshifts $z \approx \num{0.58}$ and predicted rAGNs around $z \approx \num{2.2}$. This behaviour implies that most of poorly measured sources (i.e.\ with low \texttt{band\_num} values) have similar predicted redshifts.

    Together, these results show that the broadening of distributions at $z \approx \num{0.58}$ for predicted non-AGNs and at $z \approx \num{2.2}$ for predicted AGNs correspond to sources with missing measurements that mimic some of the properties of the bulk of training sources generating a colour--redshift degeneracy. These redshifts act as sinks for faint sources that do not look as the training set and can drive the distributions of quantities derived using these values (e.g.\ linear sizes, luminosities).

\section{Linear size evolution}\label{app:linear_size}

\begin{figure}[ht!]
        \centering
        \begin{overpic}[width=0.99\linewidth, percent]{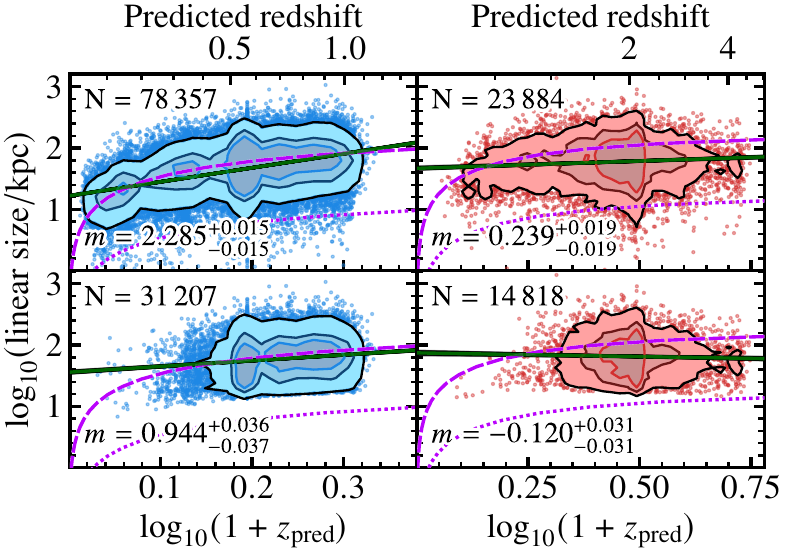}
            \put(48, 56){(a)\phantomsubcaption\label{fig:full_pred_z_vs_maj_axis_SFG}}
            \put(91, 56){(b)\phantomsubcaption\label{fig:full_pred_z_vs_maj_axis_AGN}}
            \put(48, 32){(c)\phantomsubcaption\label{fig:compl_pred_z_vs_maj_axis_SFG}}
            \put(91, 32){(d)\phantomsubcaption\label{fig:compl_pred_z_vs_maj_axis_AGN}}
        \end{overpic}
        \caption{Two-dimensional distribution of deconvolved physical size (linear size, LS) vs predicted redshift of (a) full non-AGN, (b) full AGN, (c) complete non-AGN, and (d) complete AGN samples. Description as in Fig.~\ref{fig:pred_z_vs_spec_index}. In purple, evolution of EMU-PS beam size ($\qty{18}{\arcsec}$, dashed line) and a $\qty{10}{\percent}$ of it (dotted line).}\label{fig:pred_z_vs_linear_size}
    \end{figure}
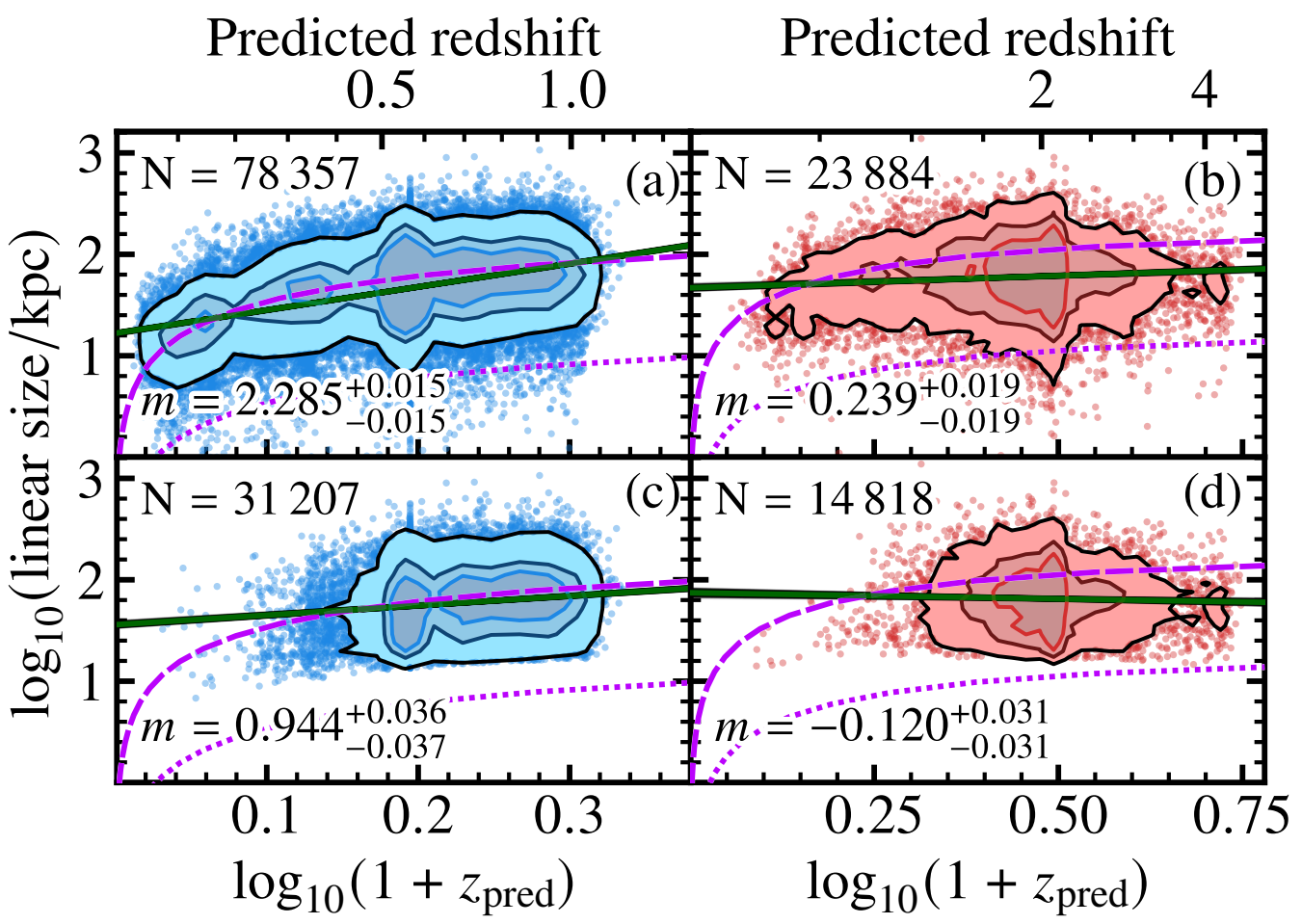

    Another dimension in the study of radio-detected sources corresponds to their physical sizes. They can help study the overall evolution of galaxies across time. Using non-negative deconvolved major axis sizes from our sample, we retrieved the distribution of projected linear sizes, LS, as a function of predicted photometric redshifts. However, before any analysis, a subset of sources must be selected to be complete in luminosity over their redshift ranges \citep[see, for instance,][]{1971MNRAS.155...95L, 2022ApJ...941...10V}. From inspecting the values in Fig.~\ref{fig:distribution_log_lum_vs_pred_z_EMU_PS}, we restricted predicted non-AGNs to $\log_{10} (L_{\qty{944}{MHz}} / \unit{\W.\Hz^{-1}}) \leq \num{23.5}$ and predicted rAGNs to $\log_{10} (L_{\qty{944}{MHz}} / \unit{\W.\Hz^{-1}}) \leq \num{25}$, calling these subsets as complete. They are shown in Fig.~\ref{fig:pred_z_vs_linear_size}.

    As with spectral indices (Sect.~\ref{sec:spectral_indices}), we fitted a line to each subset using an MCMC sampler. In the full samples, predicted redshift and LS are positively correlated, with a stronger trend for predicted non-AGNs\@. For complete subsets, results change: the best-fit slope for predicted non-AGNs is much weaker, and for complete predicted AGNs, we instead find a negative slope, consistent with previous reports of a negative evolution between photometric redshift and LS for rAGNs \citep[e.g.][]{2025A&A...700A..63K}. Negative correlations have also been found for both non-AGNs and AGNs, mainly in fainter, $\unit{\micro\jansky}$-level samples \citep[e.g.][]{2018A&A...618L...8B, 2018MNRAS.476.3544L, 2019A&A...625A.114J, 2021ApJ...910..106J}. In our case, the positive slope for complete non-AGNs might reflect an excess of very bright non-AGNs or other types of sources (e.g. HERGs, LERGs), as discussed in Sect.~\ref{sec:predicted_lums}.

    An additional analysis from Fig.~\ref{fig:pred_z_vs_linear_size} involves sources with sub-beam LS\@. Purple lines show the region where sources within $\qtylist[list-separator = {,\ \linebreak[0]}]{10;100}{\percent}$ of the synthesised beam lie. The lower limit, $\ang[angle-symbol-over-decimal]{;;1.8}$, is compatible with the value used by \citet{2021PASA...38...46N}, $\qty{2}{\arcsec}$, to separate resolved from unresolved sources. Then, all smaller sizes might not be used confidently, coinciding with our complete sample selection and the study by \citet{2012A&A...541A.135M} on sub-beam compact sources resolution.

\section{Extended EMU-PS catalogue column description}\label{app:catalogue_columns}

    We present the column description of the extended catalogue of sources from the EMU-PS included in the table \texttt{EMU\_PS\_ML}, which is available in electronic form at the CDS via \url{http://cdsweb.u-strasbg.fr/cgi-bin/qcat?J/A+A/}. Columns with the suffix \texttt{\_EMU} have been directly taken from the EMU-PS value-added catalogue of compact sources \citep{2021PASA...38...46N}. A visual representation of the sources and some of their properties is also available in the web dashboard at \url{https://racarvajalp-emups-ml-dashboard.hf.space/app}.

    \begin{itemize}[leftmargin=*, itemindent=-1em]%
        \item \texttt{Name\_CW}: Source identification as given in CatWISE\@.
        \item \texttt{RA\_CW}: Right ascension (in $\unit{\deg}$) of source in CatWISE\@.
        \item \texttt{DEC\_CW}: Declination (in $\unit{\deg}$) of source in CatWISE\@.
        \item \texttt{Name\_EMU}: Source identification as given in EMU-PS\@.
        \item \texttt{RA\_EMU}: Right ascension (in $\unit{\deg}$) of source in EMU-PS\@.
        \item \texttt{DEC\_EMU}: Declination (in $\unit{\deg}$) of source in EMU-PS\@.
        \item \texttt{sep\_CW\_EMU}: Separation (in $\unit{\asec}$) between CatWISE source and EMU-PS counterpart.
        \item \texttt{f\_peak\_EMU}: Peak EMU flux (in $\unit{\milli\jansky/\beam}$).
        \item \texttt{e\_f\_peak\_EMU}: Peak EMU flux error (in $\unit{\milli\jansky/\beam}$).
        \item \texttt{f\_int\_EMU}: Integrated EMU flux (in $\unit{\milli\jansky}$).
        \item \texttt{e\_f\_int\_EMU}: Integrated EMU flux error (in $\unit{\milli\jansky}$).
        \item \texttt{rms\_EMU}: Noise map value at EMU position (in $\unit{\milli\jansky/\beam}$).
        \item \texttt{spec\_idx\_EMU}: In-band EMU spectral index.
        \item \texttt{e\_spec\_idx\_EMU}: In-band EMU spectral index error.
        \item \texttt{maj\_dc\_EMU}: Deconvolved EMU major axis size (in $\unit{\asec}$).
        \item \texttt{e\_maj\_dc\_EMU}: Deconvolved EMU major axis size error (in $\unit{\asec}$).
        \item \texttt{gmag}: \textit{g}-band magnitude of DES-DR2 counterpart.
        \item \texttt{rmag}: \textit{r}-band magnitude of DES-DR2 counterpart.
        \item \texttt{imag}: \textit{i}-band magnitude of DES-DR2 counterpart.
        \item \texttt{zmag}: \textit{z}-band magnitude of DES-DR2 counterpart.
        \item \texttt{Ymag}: \textit{Y}-band magnitude of DES-DR2 counterpart.
        \item \texttt{Jmag}: \textit{J}-band magnitude of 2MASS counterpart.
        \item \texttt{Hmag}: \textit{H}-band magnitude of 2MASS counterpart.
        \item \texttt{Kmag}: \textit{Ks}-band magnitude of 2MASS counterpart.
        \item \texttt{W1mag}: \textit{W1}-band magnitude of source in CatWISE\@.
        \item \texttt{W2mag}: \textit{W2}-band magnitude of source in CatWISE\@.
        \item \texttt{W3mag}: \textit{W3}-band magnitude of AllWISE counterpart.
        \item \texttt{W4mag}: \textit{W4}-band magnitude of AllWISE counterpart.
        \item \texttt{band\_num}: Number of non-radio magnitudes with a detection.
        \item \texttt{class}: Original source class. \texttt{0} for non-AGNs\@. \texttt{1}, AGNs\@.
        \item \texttt{radio\_detect}: \texttt{1} if source has EMU-PS counterpart. \texttt{0} otherwise.
        \item \texttt{z\_comb}: Original source redshift.
        \item \texttt{pred\_class}: Predicted class. \texttt{0} for non-AGNs, \texttt{1} for AGNs\@.
        \item \texttt{prob\_AGN}: AGN probability, $p(\mathrm{AGN})$. Between \texttt{0} and \texttt{1}.
        \item \texttt{prob\_r\_AGN}: Radio-detectability probability for AGNs, $p(\mathrm{radio \vert AGN})$. Between \texttt{0} and \texttt{1}.
        \item \texttt{prob\_r\_nAGN}: Radio-detectability probability for non-AGNs, $p(\mathrm{radio \vert nAGN})$. Between \texttt{0} and \texttt{1}.
        \item \texttt{pred\_Z\_rAGN}: Predicted source photometric redshift if predicted as AGN\@.
        \item \texttt{pred\_Z\_rnAGN}: Predicted source photometric redshift if predicted as non-AGN\@.
        \item \texttt{pred\_Z}: Consolidated predicted source photometric redshift.
        \item \texttt{b\_pred\_Z}: Redshift prediction $\qty{68}{\percent}$ interval lower limit.
        \item \texttt{B\_pred\_Z}: Redshift prediction $\qty{68}{\percent}$ interval upper limit.
        \item \texttt{sigma\_minus}: Redshift prediction interval lower limit width.
        \item \texttt{sigma\_plus}: Redshift prediction interval upper limit width.
        \item \texttt{pred\_phys\_size}: Estimated EMU physical size (in $\unit{\kilo\parsec}$).
    \end{itemize}

\end{appendix}
\end{document}